\documentclass[12pt]{iopart}
\usepackage{graphics}
\usepackage{cite}
\begin{document}
\def\ave#1{\langle#1\rangle}
\def\Stot{S_{\rm tot}}
\def\vS{{\bi S}}
\def\Ms{M_{\rm s}}
\def\Hs{H_{\rm s}}
\def\gsim{\,$\raise0.3ex\hbox{$>$}\llap{\lower0.8ex\hbox{$\sim$}}$\,}
\def\lsim{\,$\raise0.3ex\hbox{$<$}\llap{\lower0.8ex\hbox{$\sim$}}$\,}
\jl{1}
\title
[Magnetic properties of the $S=1/2$ distorted diamond chain at $T=0$]
{Magnetic properties of the $S=1/2$ distorted diamond chain at $T=0$}
 \author{Kiyomi Okamoto\dag, Takashi Tonegawa\ddag\; and Makoto Kaburagi\S}

 \address{\dag Department of Physics,
          Tokyo Institute of Technology,
          Oh-Okayama, Meguro-ku, Tokyo 152-8551, Japan}

 \address{\ddag Department of Mechanical Engineering,
          Fukui University of Technology,
          Gakuen 3-Chome, Fukui-Shi 910-8505, Japan}

 \address{\S Faculty of Cross-Cultural Studies,
          Kobe University, Tsurukabuto, Nada-ku,
          Kobe 657-8501, Japan}
\begin{abstract}
We explore, at $T=0$, the magnetic properties of the $S=1/2$ antiferromagnetic
distorted diamond chain described by the Hamiltonian
${\cal H}
   =  \sum_{j=1}^{N/3} \left\{
         J_1 \left( {\bi S}_{3j-1} \cdot {\bi S}_{3j} 
          + {\bi S}_{3j} \cdot {\bi S}_{3j+1} \right)
       + J_2 {\bi S}_{3j+1} \cdot {\bi S}_{3j+2} 
       + J_3 \left( {\bi S}_{3j-2} \cdot {\bi S}_{3j} 
          + {\bi S}_{3j} \cdot {\bi S}_{3j+2} \right)
     \right\} \allowbreak
    - H \sum_{l=1}^{N} S_l^z
$
with $J_1,\;J_2,\;J_3\ge0$, which well models 
${\rm A_3 Cu_3 (PO_4)_4}$ with ${\rm A = Ca, \,Sr}$,
${\rm Bi_4 Cu_3 V_2 O_{14}}$
and azurite $\rm Cu_3(OH)_2(CO_3)_2$.
We employ the physical consideration, the
degenerate perturbation theory, the level spectroscopy analysis of the
numerical diagonalization data obtained by the Lanczos method and also the
density matrix renormalization group (DMRG) method.  We investigate the
mechanisms of the magnetization plateaux at $M=\Ms/3$ and $M=(2/3)\Ms$,
and also show the precise phase diagrams on the $(J_2/J_1,\;J_3/J_1)$ plane
concerning with these magnetization plateaux, where $M=\sum_{l=1}^{N} S_l^z$
and $\Ms$ is the saturation
magnetization.  We also calculate the magnetization curves and the
magnetization phase diagrams by means of the DMRG method.
\end{abstract}

\maketitle
%
%
%********************************************
\section{Introduction}
%********************************************
%
Low-dimensional quantum spin systems have attracted increasing attention in
recent years.  A few years ago, Ishii, Tanaka, Mori, Uekusa, Ohashi, Tatani,
Narumi and Kindo \cite{Ishii} have reported the experimental results on a
trimerized $S=1/2$ quantum spin
chain $\rm Cu_3 Cl_6 (H_2 O)_2 \cdot 2H_8 C_4 SO_2$.  From the
structure-analysis experiment they have proposed a model shown in figure 1
for this substance \cite{Ishii}.  The Hamiltonian of this model is written as
\begin{eqnarray}
  &&{\cal H} = {\cal H}_0 + {\cal H}_{\rm Z} \\
  &&{\cal H}_0
   =  J_1 \sum_{j=1}^{N/3} \left( {\bi S}_{3j-1} \cdot {\bi S}_{3j} 
          + {\bi S}_{3j} \cdot {\bi S}_{3j+1} \right)
    + J_2 \sum_{j=1}^{N/3} {\bi S}_{3j+1} \cdot {\bi S}_{3j+2} \nonumber \\
  &&~~~~~~~~~~~ + J_3 \sum_{j=1}^{N/3} \left( {\bi S}_{3j-2} \cdot {\bi S}_{3j}
          + {\bi S}_{3j} \cdot {\bi S}_{3j+2} \right) \\
  &&{\cal H}_{\rm Z}
   = -H \sum_{l=1}^{N} S_l^z.
   \label{eq:ham}
\end{eqnarray}
where ${\bi S}_{l}$ is the $S=1/2$ operator at the $l$th site and $N$
is the total number of spins in the system.  All the coupling constants, $J_1$,
$J_2$ and $J_3$, are supposed to be positive (antiferromagnetic).  Hereafter
we set $\tilde J_2 \equiv J_2/J_1$ and $\tilde J_3 \equiv J_3/J_1$.  The
magnetic field is denoted by $H$.  We can assume $J_1\ge J_3$ without loss of
generality because the $J_1<J_3$ case is equivalent to the $J_1>J_3$ case by
interchanging the role of $J_1$ and $J_3$.

\begin{figure}[ht]
   \begin{center}
      \scalebox{0.5}[0.5]{\includegraphics{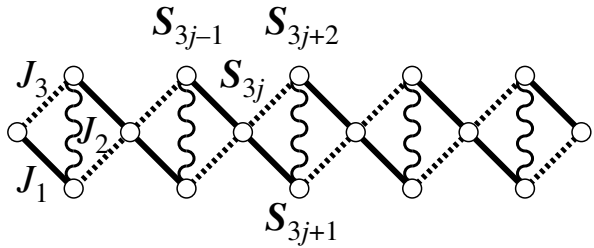}}
   \end{center}
   \caption{Sketch of the distorted diamond (DD) chain.
            Solid lines denote the coupling $J_1$,
            wavy lines the coupling $J_2$,
            and dotted lines the coupling $J_3$.
            We can assume $J_1 \ge J_3$ without loss of generality.}
   \label{fig:model}
\end{figure}

This model has been named the \lq\lq distorted diamond (DD) chain
model\rq\rq \cite{Oka-Tone}.  The name \lq\lq diamond\rq\rq\ comes from the
mark of the playing cards, and \lq\lq distorted\rq\rq\ from the fact $J_1\ne J_3$ in
general. The symmetric case ($J_1=J_3$) has first been proposed by Takano,
Kubo and Sakamoto (TKS) \cite{TKS} earlier than Ishii et al. \cite{Ishii} and
ourselves \cite{Oka-Tone}.  We \cite{Oka-Tone} have investigated the ground
state properties of the $S=1/2$ DD chain model when $H=0$ by an analytical
method as well as the level spectroscopy (LS) analysis of the numerical
diagonalization data obtained by the Lanczos technique.  Later the theoretical
studies on the $S=1/2$ DD chain have also been reported by ourselves
\cite{Tone-Oka-1,Tone-Oka-2}, Sano and Takano \cite{Sano-Takano} and Honecker
and L\"auchli \cite{Honecker-Lauchli}.

Unfortunately the substance which was thought to be
$\rm Cu_3 Cl_6 (H_2 O)_2 \cdot 2H_8 C_4 SO_2$ at first \cite{Ishii}
has been proved to be $\rm Cu_2 Cl_4 \cdot H_8 C_4 SO_2$, the lattice of
which is the two-leg zig-zag chain with the bond-alternation by the same
group \cite{Fujisawa}. 
Now, however, the $S=1/2$ DD chain model is again in
spotlight as a model for
${\rm A_3 Cu_3 (PO_4)_4}$ with ${\rm A = Ca, \,Sr}$
\cite{Drillon1, Drillon2, Ajiro},
${\rm Bi_4 Cu_3 V_2 O_{14}}$ \cite{Sakurai}
and azurite $\rm Cu_3(OH)_2(CO_3)_2$.

For ${\rm A_3 Cu_3 (PO_4)_4}$ with ${\rm A = Ca, \,Sr}$,
Drillon et al. \cite{Drillon1,Drillon2} measured
the magnetic susceptibility $\chi(T)$, the specific heat $C(T)$
and the magnetization process $M(H)$,
and proposed the model without $J_2$ interactions.
Ajiro et al. \cite{Ajiro} performed the high field magnetization process
measurement up to 40T and the neutron diffraction experiment,
and proposed the model with $J_2$ interactions.
In these reports,
a wide magnetization plateau at $M=\Ms/3$ was observed,
where $M$ is the $z$ component of the total spin,
defined by $M=\sum_{l=1}^N S_l^z$, and
$\Ms$ is the saturation magnetization.
The behavior of $\chi(T)$ and neutron diffraction results
suggest that these substances are ferrimagnetic
above the three-dimensional ordering temperature.

Sakurai et al. \cite{Sakurai}
reported $\chi(T)$, $M(H)$ (up to 28T), $C(T)$ and the ${\rm {}^{51}V}$ NMR
of ${\rm Bi_4 Cu_3 V_2 O_{14}}$.
Unfortunately,
they could not obtain any conclusion on the existence of the $M=\Ms/3$ plateau
because $M \sim 0.27\Ms$ even when $H=28\,{\rm T}$.
The ferrimagnetic behavior was not seen in $\chi(T)$
above the three-dimensional ordering temperature. 
It seems that this substance has the spin-fluid (SF) ground state
above the three-dimensional ordering temperature.

Very recently, 
Kikuchi and co-workers \cite{Kikuchi1, Kikuchi2, Kikuchi3} have
reported the experimental results on the magnetic and thermal properties of
azurite $\rm Cu_3 (OH)_2 (CO_3)_2$.  
Following their reports, azurite has the spin-fluid (SF)
ground state and the wide magnetization plateau at $M=\Ms/3$.

Thus, above the three-dimensional ordering temperature,
the ground states of ${\rm A_3 Cu_3 (PO_4)_4}$ with ${\rm A = Ca, \,Sr}$
seem to be ferrimagnetic,
whereas those of ${\rm Bi_4 Cu_3 V_2 O_{14}}$
and azurite $\rm Cu_3 (OH)_2 (CO_3)_2$ seem to be SF.
In view of these situations,
we think it is very important
to investigate the magnetic properties of the $S=1/2$ DD chain model in more
detail.  
Throughout this paper we consider the $T=0$ case.

This paper is organized as follows.  The ground-state properties at zero
magnetic field are reviewed in \S2.  The magnetization plateaux at
$M=\Ms/3$ and $M=(2/3)\Ms$ are discussed in \S3 and \S4, respectively.  The
numerical results for the magnetization curves and the magnetization phase
diagram are presented in \S5.  The last section \S6 is devoted to discussion
and summary.

%********************************************
\section{Review of the ground-state properties at zero magnetic field}
%********************************************

\begin{figure}[ht]
   \begin{center}
         \scalebox{0.35}[0.35]{\includegraphics{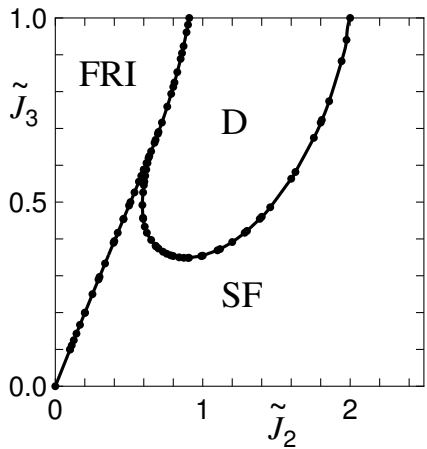}}
   \end{center}
   \caption{Zero field ground state phase diagram of the $S=1/2$ DD chain
            model.  The $\tilde J_3=1$ case is reduced to the model of
            TKS \cite{TKS}.
            There are three phase in this phase diagram;
            the ferrimagnetic (FRI) phase, the dimer (D) phase and the
            spin-fluid (SF) phase.}
   \label{fig:phase-m0}
\end{figure}

Discussions presented in this section are confined to the case where
$H=0$.  We \cite{Oka-Tone} have investigated the ground state of the $S=1/2$
DD chain model by an analytical method as well as the LS analysis of the
numerical diagonalization data obtained by the Lanczos technique, and obtained 
the phase diagram shown in figure \ref{fig:phase-m0}.  There are three phases
in this phase diagram; the ferrimagnetic (FRI) phase, the dimer (D) phase and
the spin-fluid (SF) phase.  The magnitude $\Stot$ of the total spin
$\vS_{\rm tot}$, defined by $\vS_{\rm tot}\equiv\sum_{l=1}^N\vS_l$ and
$\vS_{\rm tot}^2=\Stot(\Stot+1)$, is $\Stot = N/6$ in the FRI phase, whereas
$\Stot=0$ in the D and SF phases.  In the D phase, there is a finite energy
gap between the doubly degenerate ground state and the first excited state,
while in the SF phase there is no energy gap.

\begin{figure}[ht]
   \begin{center}
         (a) \scalebox{0.35}[0.35]{\includegraphics{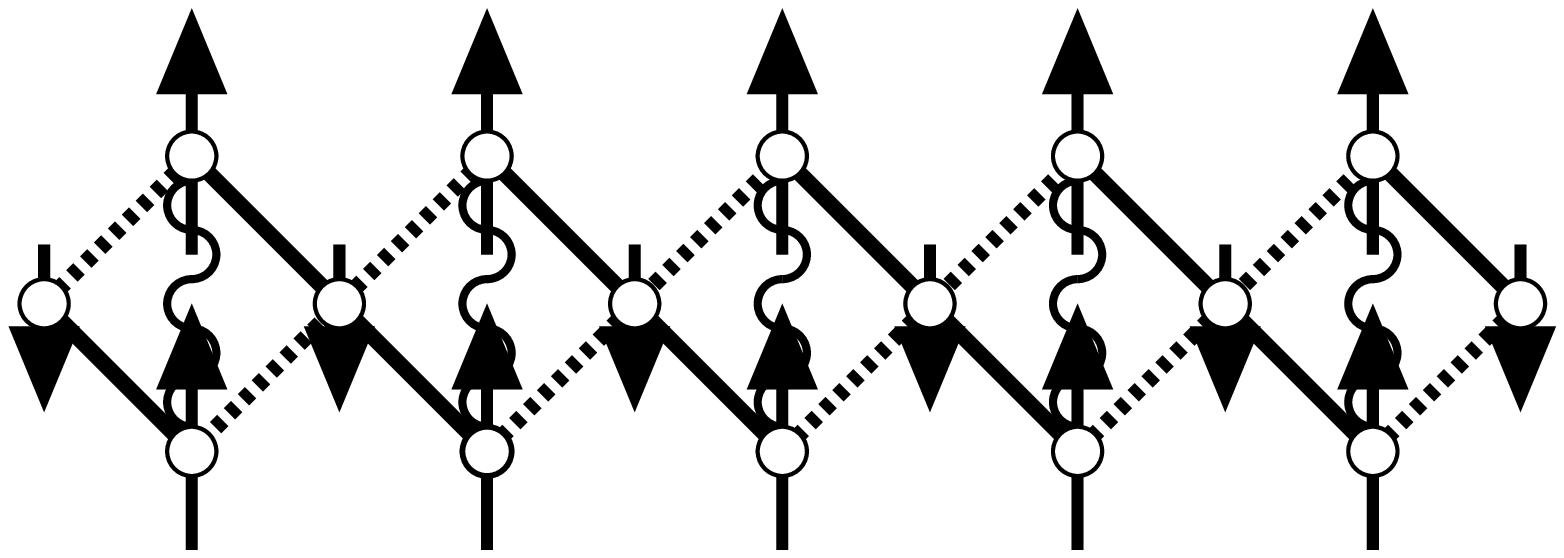}}~~~
         (b) \scalebox{0.35}[0.35]{\includegraphics{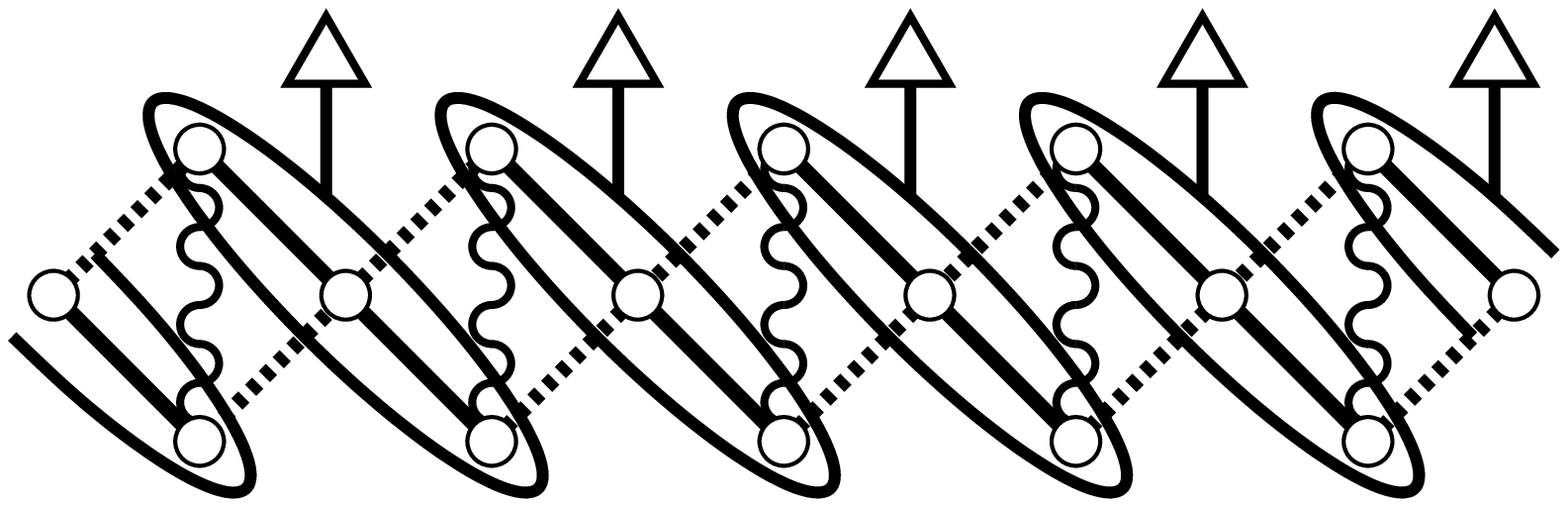}}
   \end{center}
   \caption{Physical pictures of the FRI ground state for the $S=1/2$ DD
            chain model.
            Open ellipses with open triangles in (b) denote the 3-spin
            clusters in the state $\phi_1$.}
   \label{fig:ferri-picture}
\end{figure}

When $J_2=0$, we can readily know that the ground state is the FRI state
with $S_{\rm tot}=N/6$ by use of the Lieb-Mattis theorem \cite{L-M}.  The
physical pictures of the FRI state are shown in
figure \ref{fig:ferri-picture}, where the picture (a) holds good for the
$J_1\sim J_3$ case, whereas (b) for the $J_1\gg J_2,\;J_3$ case.  The ellipses
in (b) denote the effective 3-spin cluster which is formed when the
interaction $J_1$ is much larger than the interactions
$J_2$ and $J_3$.  We represent the $S_l^z=1/2$ and $S_l^z=-1/2$ states of the
$l$th single spin by $\uparrow_l$ and $\downarrow_l$, respectively.  Then,
the ground state wave functions of the $j$th cluster in the case of
$J_2=J_3=0$ are expressed as
\begin{equation}
  \phi_{1,j}
    = {1 \over \sqrt{6}}
    \left(   |\uparrow_{3j-1}\uparrow_{3j}\downarrow_{3j+1}\rangle
          - 2|\uparrow_{3j-1}\downarrow_{3j}\uparrow_{3j+1}\rangle
          +  |\downarrow_{3j-1}\uparrow_{3j}\uparrow_{3j+1}\rangle
    \right)
    \label{eq:sz+12}
\end{equation}
and
\begin{equation}
  \phi_{2,j}
    = {1 \over \sqrt{6}}
    \left(   |\downarrow_{3j-1}\downarrow_{3j}\uparrow_{3j+1}\rangle
          - 2|\downarrow_{3j-1}\uparrow_{3j}\downarrow_{3j+1}\rangle
          +  |\uparrow_{3j-1}\downarrow_{3j}\downarrow_{3j+1}\rangle
    \right)
    \label{eq:sz-12}
\end{equation}
for $S_{\rm tot}^{(3)z}=1/2$ and $S_{\rm tot}^{(3)z}=-1/2$, respectively,
where $S_{\rm tot}^{(3)z}$ is the $z$ component of the total spin of the
3-spin cluster.  When all the 3-spin clusters are effectively in the $\phi_1$
(or $\phi_2$) state, the FRI state shown by figure 3(b) is realized.  This
physical consideration has been developed in
\cite{Tone-Oka-1,Honecker-Lauchli} by use of the degenerate perturbation
theory \cite{Totsuka} around the point $\tilde J_2 = \tilde J_3 =0$.
In fact, the quantum phase transition between the FRI
and SF phases takes place at $\tilde J_3=\tilde J_2$ near
$(\tilde J_2,\tilde J_3)=(0,0)$, and it is of the first order.  This shows
very good agreement with the numerical results (see figure
\ref{fig:phase-m0}).  The quantum phase transition between the FRI and D
phases is also of the first order.

In the case of $J_3=0$, our DD chain is reduced to the $J_1$-$J_1$-$J_2$
trimerized antiferromagnetic chain, the ground state of which is the SF
state.  In particular, in the case of $J_1=J_2$, it is further simplified to
the uniform antiferromagnetic chain.

\begin{figure}[ht]
   \begin{center}
     (a) \scalebox{0.35}[0.35]{\includegraphics{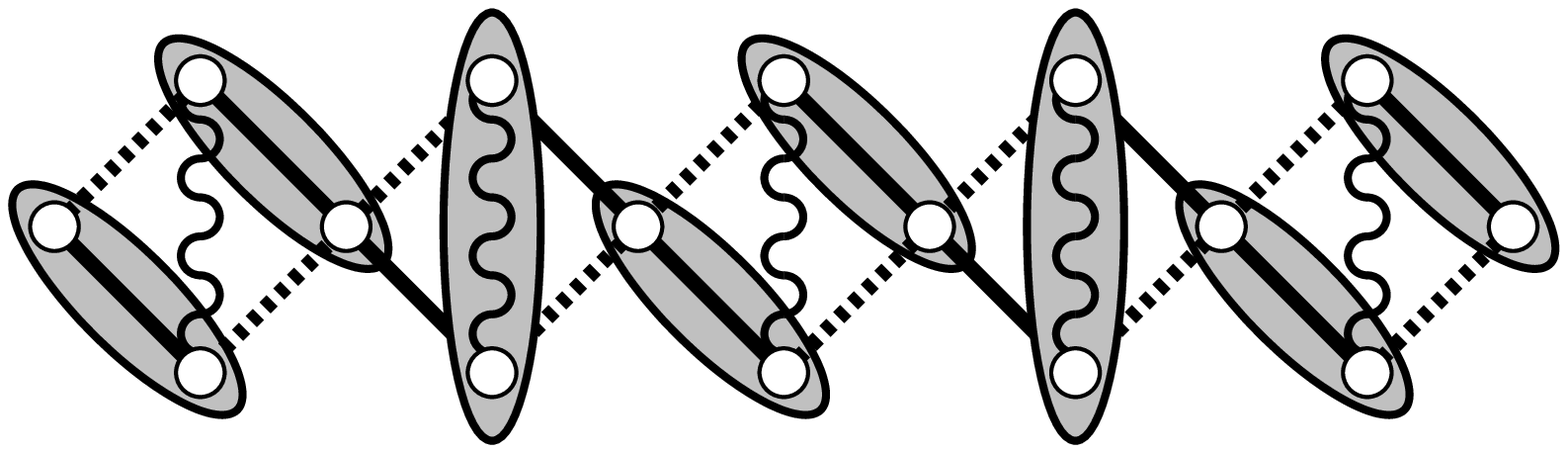}}~~~
     (b) \scalebox{0.35}[0.35]{\includegraphics{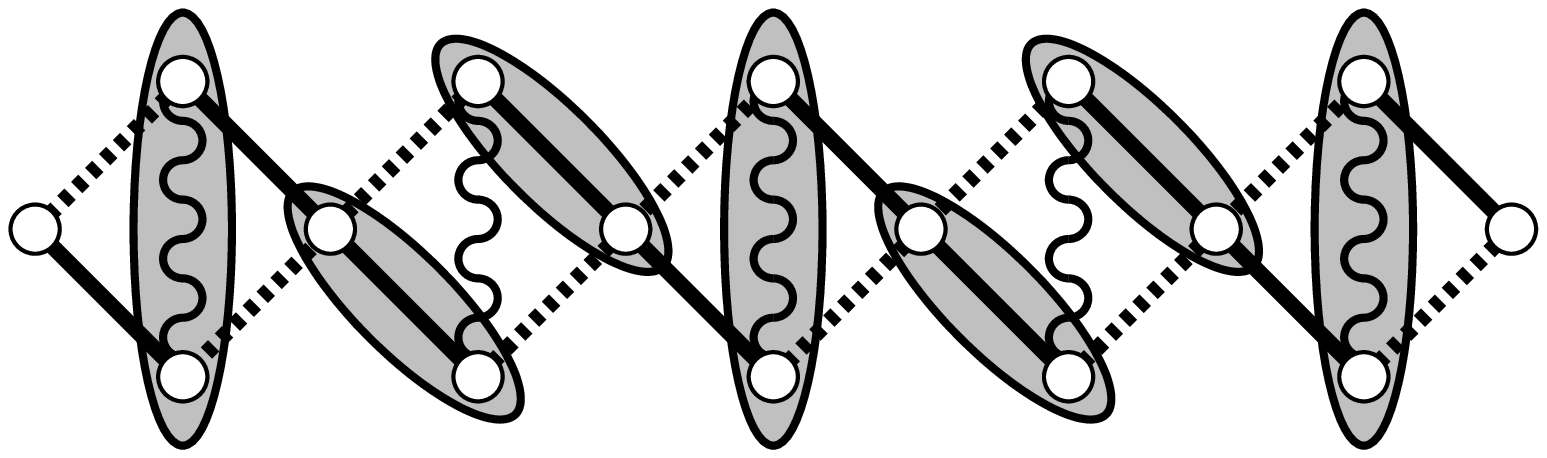}}
   \end{center}
   \caption{Physical pictures of the D ground state for the $S=1/2$ DD chain
            model.
            Two spins in a shadowed ellipse effectively form a singlet dimer pair.}
   \label{fig:dimer}
\end{figure}

The D phase is caused by the frustration, and is doubly degenerate due to
the spontaneous breaking of the translational symmetry (SBTS), as shown in
figure \ref{fig:dimer}.  The D phase is realized in the region where the
frustration effect seems to be severe.  We note that there is no frustration
when $J_2=0$ or $J_3=0$.  The quantum phase transition between the SF and  D
phases is essentially the same as that in the $S=1/2$ chain with
next-nearest-neighbor interactions \cite{Tone-Hara,Oka-Nom}, and is of the
Berezinskii-Kosterlitz-Thouless (BKT) type \cite{Berezinskii,KT}, as has been
fully discussed in our previous paper \cite{Oka-Tone}.
\begin{figure}[h]
   \begin{center}
       \hbox{~~~~~~~~~~~~~~~~~\Large (a)}
       \scalebox{0.5}[0.5]{\includegraphics{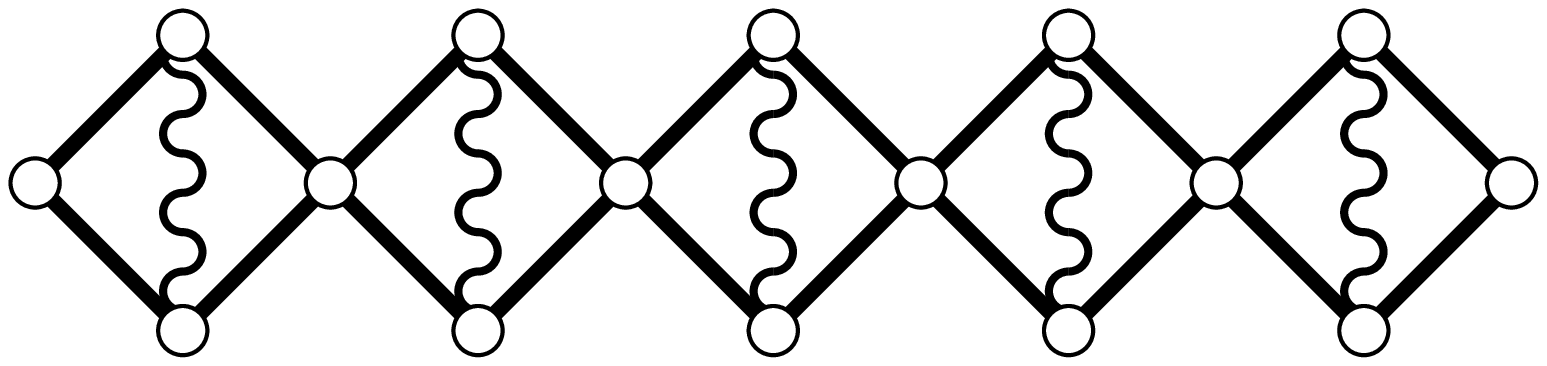}}
    \end{center}
   \begin{center}
       \hbox{~~~~~~~~~~~~~~~~~\Large (b)}
       \scalebox{0.5}[0.5]{\includegraphics{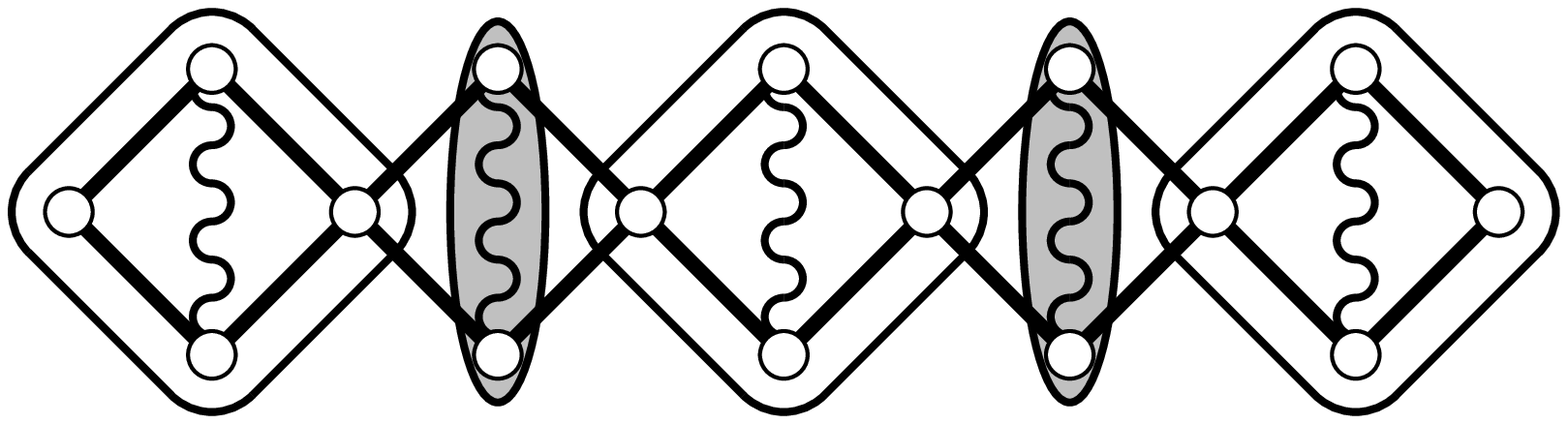}}
   \end{center}
   \begin{center}
       \hbox{~~~~~~~~~~~~~~~~~\Large (c)}
       \scalebox{0.5}[0.5]{\includegraphics{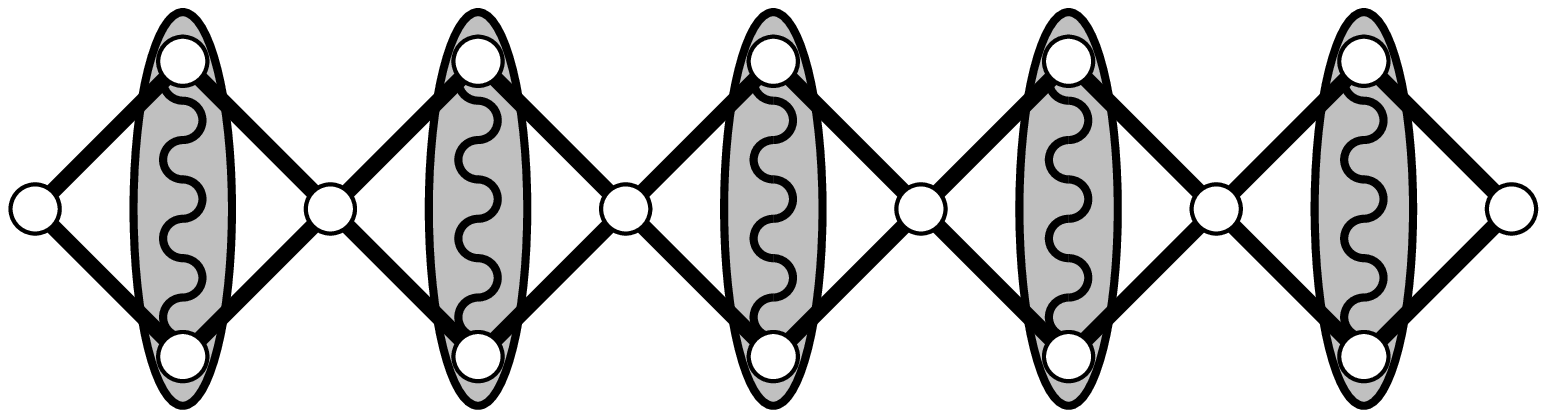}}
   \end{center}
   \caption{(a) The model of Takano, Kubo and Sakamoto \cite{TKS}.
            (b) The tetramer-dimer (TD) state, where the rectangles
            represent tetramers and the ellipses singlet dimers.
            (c) The dimer-monomer (DM) state.}
   \label{fig:TKS}
\end{figure}

As we have mentioned in \S1, the symmetric case $J_1=J_3$, which is shown in
figure \ref{fig:TKS}(a), has first been proposed by TKS \cite{TKS}.  They
have found that the ground state is the FRI state, the tetramer-dimer (TD)
state or the dimer-monomer (DM) state, depending on whether
$\tilde J_2<0.909$, $0.909<\tilde J_2 <2$ or $2<\tilde J_2$.  The value
$0.909$ of the FRI-TD boundary has been calculated by the numerical method,
whereas the value $2$ of the TD-DM boundary is the exact one obtained by an
analytical method.  In the TD state, as shown in figure \ref{fig:TKS}(b),
four spins of the diamond unit ${\bi S}_{3j-3}$, ${\bi S}_{3j-2}$,
${\bi S}_{3j-1}$ and ${\bi S}_{3j}$ (or ${\bi S}_{3j}$, ${\bi S}_{3j+1}$,
${\bi S}_{3j+2}$ and ${\bi S}_{3j+3}$) form a tetramer, and two spins
${\bi S}_{3j+1}$ and ${\bi S}_{3j+2}$ (or ${\bi S}_{3j+4}$ and
${\bi S}_{3j+5}$) form a singlet dimer.  Obviously, the TD state is doubly
degenerate due to the SBTS.  In the DM state, on the other hand, two spins
${\bi S}_{3j+1}$ and ${\bi S}_{3j+2}$, connected by the $J_2$ interactions,
form a singlet dimer and the remaining joint spins ${\bi S}_{3j}$ are
completely free, as shown in figure \ref{fig:TKS}(c).  Thus, the DM state is
$2^{N/3}$-fold degenerate.  It is very interesting that the ground state can
be exactly expressed as the direct product of the local states in a finite
range of the parameter region $\tilde J_2>0.909$.

The TD state is very peculiar to the $J_1 = J_3$ case.  Namely, if $J_1>J_3$,
the tetramer which consists of ${\bi S}_{3j-3}$, ${\bi S}_{3j-2}$,
${\bi S}_{3j-1}$ and ${\bi S}_{3j}$ is effectively decomposed into two dimers,
the $({\bi S}_{3j-3}$, ${\bi S}_{3j-2})$ pair and the
$({\bi S}_{3j-1}$, ${\bi S}_{3j})$ pair.  Two spins in each pair are connected
by the $J_1$ interaction which is larger than the $J_3$ one.  This
decomposition leads to the doubly degenerate D state shown in
figure \ref{fig:dimer}.  Thus, the TD state is the special case of the doubly
degenerate D state.

The $2^{N/3}$-fold degeneracy of the DM state is lifted when
$J_1\ne J_3$.  In other words, the effective interaction between the free
spins ${\bi S}_{3j}$ and ${\bi S}_{3(j+1)}$ appears through the dimer located
between them.  Honecker and L\"auchli \cite{Honecker-Lauchli} have
derived the effective interaction between ${\bi S}_{3j}$ and
${\bi S}_{3(j+1)}$ as
\begin{equation}
    J_{\rm eff}
    = {(J_1 - J_3)^2 \over J_2} \left( {1 \over 2} + \cdots  \right)
    \label{eq:Jeff}
\end{equation}
which leads to the SF ground state.  The effective interaction $J_{\rm eff}$
vanishes when $J_1=J_3$, which is consistent with TKS's result \cite{TKS}
on the DM state. Then, the DM state is also peculiar to the $J_1=J_3$ case
and is the special case of the SF state.

%********************************************
\section{Magnetization plateau at $M=\Ms/3$}
%********************************************

\begin{figure}[ht]
   \begin{center}
     \scalebox{0.5}[0.5]{\includegraphics{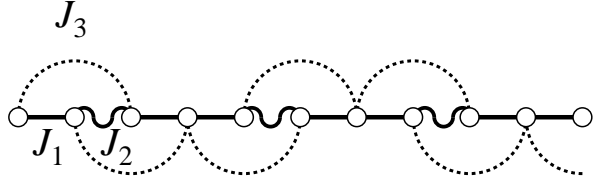}}
   \end{center}
   \caption{Single chain representation of the DD chain model.}
   \label{fig:chain-rep}
\end{figure}

\begin{figure}[ht]
   \begin{center}
         \scalebox{0.35}[0.35]{\includegraphics{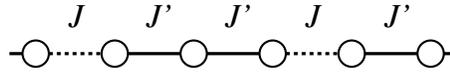}}
   \end{center}
   \caption{Antiferromagnetic chain with the trimerization.}
   \label{fig:trimer}
\end{figure}

In this section, we discuss the magnetization plateau at $M=\Ms/3$.  Since
the Hamiltonian has the trimer nature, as is shown in
figure \ref{fig:chain-rep}, the mechanism for the $M=\Ms/3$ plateau is
just the same as that for the trimerized chain (Figure \ref{fig:trimer})
investigated by Okamoto and Kitazawa \cite{Oka-Kita} and by
Honecker \cite{Honecker99}.  The $M=\Ms/3$ plateau can be realized without
any SBTS.  This is consistent with the necessary condition for the
magnetization plateau by Oshikawa, Yamanaka and Affleck (OYA) \cite{OYA},
\begin{equation}
  n (S - \langle m \rangle) = {\rm integer}
  \label{eq:OYA}
\end{equation}
where $n$ is the periodicity of the wave function of the plateau state, $S$
the magnitude of spins and $\langle m \rangle$ the average magnetization per
one spin in the plateau.

\begin{figure}[ht]
   \begin{center}
         (a)
         \scalebox{0.35}[0.35]{\includegraphics{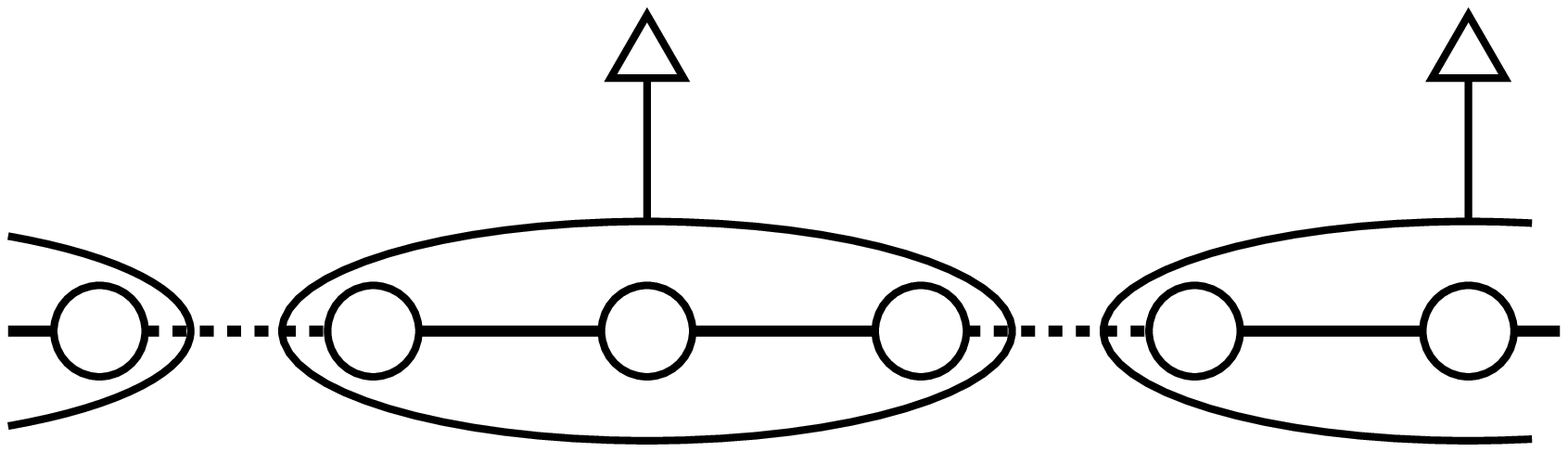}}~~~
         (b)
         \scalebox{0.35}[0.35]{\includegraphics{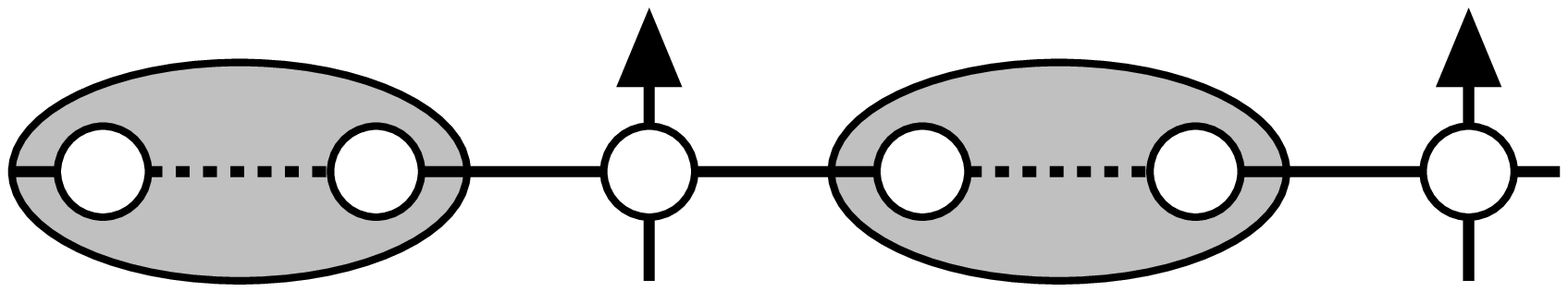}}
   \end{center}
   \caption{Two mechanisms for the $M=\Ms/3$ plateau for the $S=1/2$
            trimerized chain: (a) the plateau A and (b) the plateau B.
            Open ellipses denote the trimers with $S^{(3)z}_{\rm tot}=1/2$
            and shaded ellipses denote the singlet dimers}
   \label{fig:pl-mech0}
\end{figure}

Here we explain the mechanism for the $M=\Ms/3$ plateau of the $S=1/2$
trimerized chain.    When $0<J\ll J'$, each of three spins
connected by $J'$ forms an effective trimer with $S^{(3)z}_{\rm tot}=1/2$ (or
$-1/2$), the wave function of which is $\phi_1$ (or $\phi_2$).  When the
magnetic field is applied, all the trimers belong to the
$S^{(3)z}_{\rm tot}=1/2$ state, which brings about $M=\Ms/3$, as shown
in figure \ref{fig:pl-mech0}(a). The plateau due to this mechanism is named
the \lq\lq plateau A".  The mechanism of the plateau A state is essentially
the same as that of the FRI state for the $J_1\gg J_2,\;J_3$ case when
$H=0$, which is already explained in \S2.  
When $J\gg J'>0$, on the other hand, each pair of two spins connected by $J$
forms an effective singlet dimer pair and the remaining spins are nearly
free.  When the magnetic field is applied, the nearly free spins turn to the
direction of the field, resulting also in $M=\Ms/3$, as shown in
figure \ref{fig:pl-mech0}(b).  We call the plateau due to this mechanism the
\lq\lq plateau B".
Thus, there are two mechanisms for the $M=\Ms/3$ plateau.  It is
easily seen that the change of the plateau mechanism occurs at $J=J'$, where
the trimerized chain is reduced to the uniform chain having no gap (i.e., no
plateau) in the excitation spectrum.  This situation is most simply seen in
the trimerized $XY$ chain \cite{Okamoto-SSC92} which is exactly solvable.

\begin{figure}[ht]
   \begin{center}
         (a)
         \scalebox{0.35}[0.35]{\includegraphics{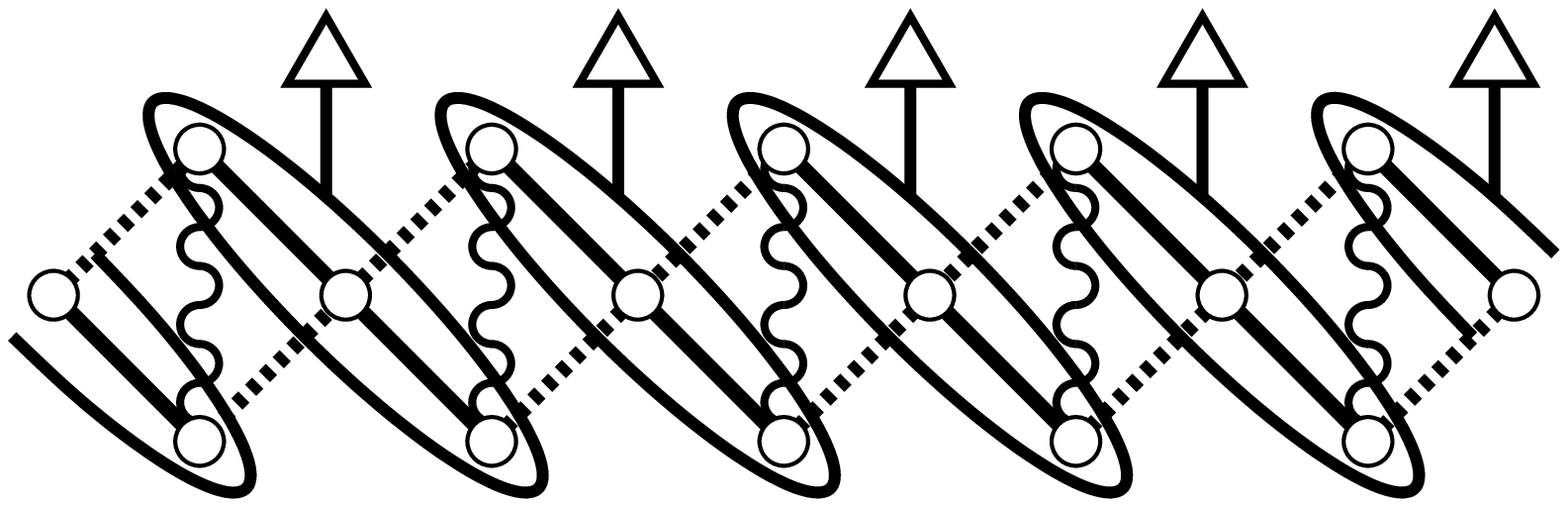}}
         (b)
         \scalebox{0.35}[0.35]{\includegraphics{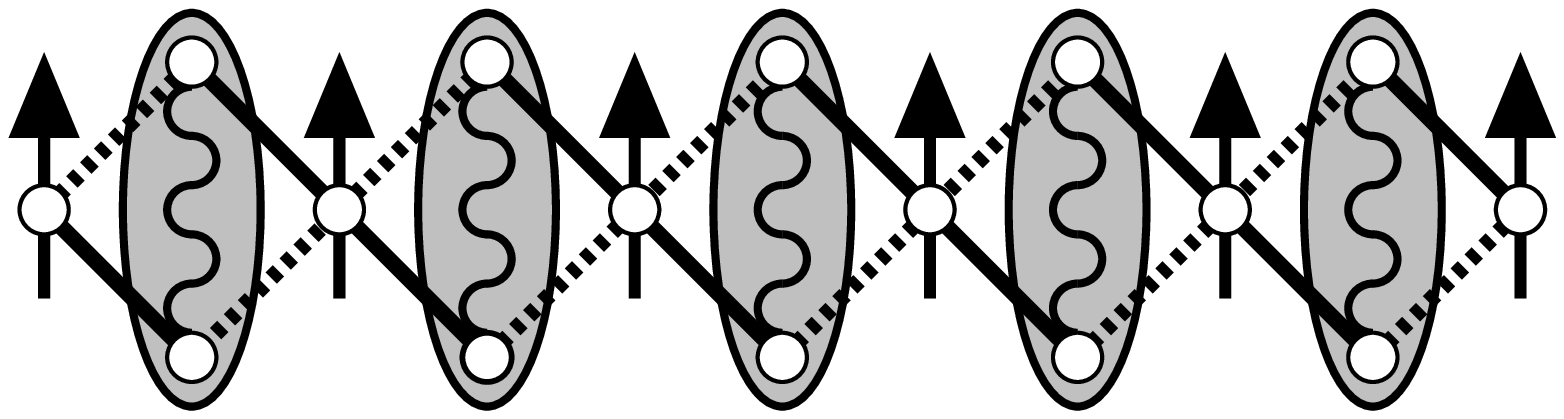}}
   \end{center}
   \caption{Two mechanisms for the $M=\Ms/3$ plateau for the $S=1/2$ DD
            chain: (a) the plateau A and (b) the plateau B.
            Shaded ellipses denote the singlet dimers
            and open ellipses denote the trimers with
            $S_{\rm tot}^{(3)z} = 1/2$.}
   \label{fig:pl-mech}
\end{figure}

A similar situation is expected to hold for the present $S=1/2$ DD chain
model.  The pictures of the plateaux A and B are depicted in figure
\ref{fig:pl-mech}(a) and (b), respectively, by use of the diamond form.  When
$J_3=0$, our model is reduced to the simple trimerized chain discussed
above.  Then, the change of the plateau mechanism occurs at $J_1=J_2$ when
$J_3=0$.  To find the phase boundary between the plateaux A and B for the
general case, we have to rely on the numerical method.  One of the most
powerful methods is the level spectroscopy (LS) \cite{Oka-LS, Nom-LS}.
Kitazawa \cite{Kitazawa}
has developed the LS for this kind of transition (belonging to the Gaussian
universality class) by use of the twisted boundary condition (TBC) instead of
the periodic boundary condition (PBC).  This method has been applied to
several magnetization plateau problems in a successful way \cite{Oka-Kita,
Kita-Oka1, Kita-Oka2, Oka-Sakai}.

In finite size systems with the PBC, the $M=\Ms/3$ plateau state changes
smoothly from the plateau A state to the plateau B state when the parameter
$J_2/J_1$ runs from 0 to $\infty$, although the plateau width takes the
smallest value at the transition point.  Under the TBC
\begin{equation}
    {\bi S}_N \cdot {\bi S}_1
    \Rightarrow - S_N^x S_1^x - S_N^y S_1^y + S_N^z S_1^z,
\end{equation}
on the other hand, the plateau A state and the plateau B state can be easily
distinguished even in finite size systems by the eigenvalue $P$ of the space
inversion operator
\begin{equation}
   {\bi S}_j \Rightarrow {\bi S}_{N-j+1}.
\end{equation}
Since it is believed that the boundary condition does not affect any
physical quantities in the $N\to\infty$ limit, we distinguish the ground
state by comparing the lowest energies with $P=\pm 1$.  In other words, we
can find the phase boundary between the plateaux A and B from the crossing of
the lowest energies with $P=\pm 1$ as functions of the quantum parameters in
the Hamiltonian.  This is the physical interpretation of Kitazawa's TBC
method \cite{Kitazawa}.  He \cite{Kitazawa} has consolidated the foundations
of his TBC method by use of the bosonization, the renormalization group
method and the conformal field theory.

\begin{figure}[ht]
   \begin{center}
         \scalebox{0.4}[0.4]{\includegraphics{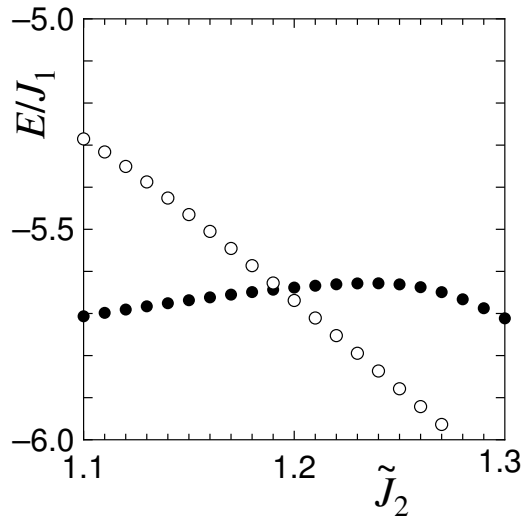}}
   \end{center}
   \caption{Crossing of the lowest energies in the $M=\Ms/3$
            sector with $P=1$ (closed circles) and $P=-1$ (open circles)
            for the $N=18$ and $\tilde J_3=0.55$ case under the TBC,
            when $\tilde J_2$ is run.
            The crossing point of these two curves
            locates at $\tilde J_2 \simeq 1.193$.}
   \label{fig:cross}
\end{figure}

\begin{figure}[ht]
   \begin{center}
         \scalebox{0.4}[0.4]{\includegraphics{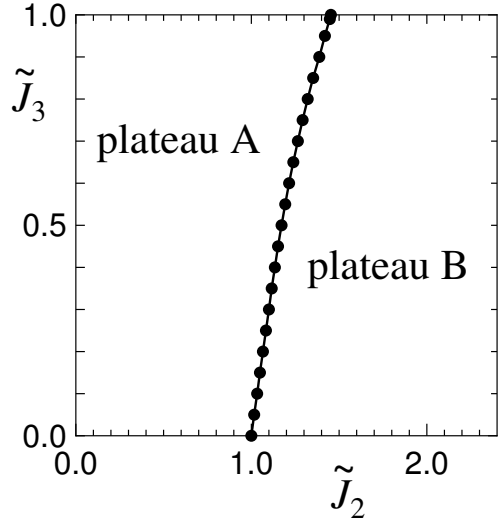}}
   \end{center}
   \caption{$M=\Ms/3$ plateau phase diagram of the $S=1/2$ DD chain model.}
   \label{fig:phase-1-3}
\end{figure}

Figure \ref{fig:cross} shows the crossing of the lowest energies with
$P=\pm 1$ when $N=18$ and $\tilde J_3=0.55$.  We
can see that two curves cross with each other at
$\tilde J_2^{\rm (cr)}(18) \simeq 1.193$, close to which the quantum
transition point $\tilde J_2^{\rm (cr)}$ between the plateau A and B phases
for the $N\to\infty$ system is expected to be located.  An accurate value of
$\tilde J_2^{\rm (cr)}$ may be estimated by extrapolating
$\tilde J_2^{\rm (cr)}(N)$ to the $N\to\infty$ limit.  Performing this
extrapolation we have assumed an $N$ dependence of $\tilde J_2^{\rm (cr)}(N)$
as
\begin{equation}
  \tilde J_2^{\rm (cr)}(N)
     = \tilde J_2^{\rm (cr)} + c_1/N^2 + c_2/N^4
   \label{eq:extrapolation}
\end{equation}
where $c_1$ and $c_2$ are numerical constants, and have employed the finite
size values, $\tilde J_2^{\rm (cr)}(12)$, $\tilde J_2^{\rm (cr)}(18)$ and
$\tilde J_2^{\rm (cr)}(24)$.  Repeating this procedure with sweeping
$\tilde J_3$, we have obtained the $M=\Ms/3$ plateau phase diagram shown in
figure \ref{fig:phase-1-3}.  The estimated value of errors of the phase
boundary is, for instance, $\tilde J_2^{\rm (cr)}=1.193 \pm 0.001$ (or better)
for $\tilde J_3=0.55$.

\begin{figure}[ht]
   \begin{center}
        \scalebox{0.4}[0.4]{\includegraphics{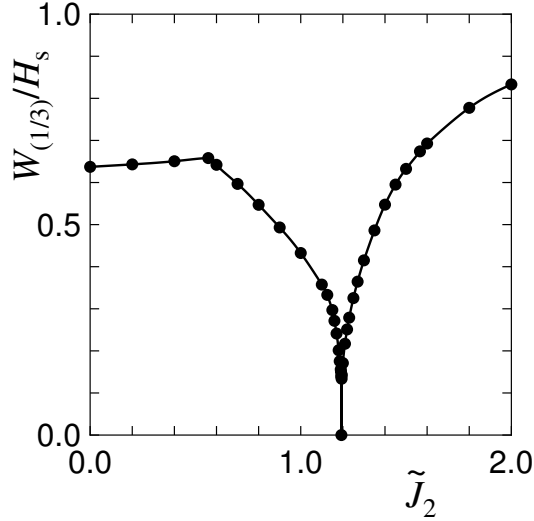}}
   \end{center}
   \caption{Normalized $M=\Ms/3$ plateau width $W_{(1/3)}/\Hs$ as a function
   of $\tilde J_2$ when $\tilde J_3=0.55$,
   where $\Hs$ is the saturation field.
   The plateau width $W_{(1/3)}$ vanishes at $\tilde J_3 \simeq 1.193$,
   which well agrees with the transition point between the plateaux A and B
   in figure \ref{fig:phase-1-3}.
   The cusp at $\tilde J_3 \simeq 0.561$ corresponds to the 
   FRI-SF transition in figure \ref{fig:phase-m0}.}
   \label{fig:width-1-3}
\end{figure}

We have performed the density matrix renormalization group (DMRG) calculation
\cite{White1,White2} to obtain the ground state magnetization curves 
and the magnetization phase diagrams for our $S=1/2$ DD chain model, 
the details of which will be described in \S5.  
Here, we present in figure \ref{fig:width-1-3} 
the width $W_{(1/3)}$ (see equation (29) below) of the $M=\Ms/3$ plateau
for $\tilde J_3 = 0.55$ in the $N \to \infty$ limit;
in this figure $W_{(1/3)}$ is plotted as a function of
$\tilde J_2$.  The point where $W_{(1/3)}$ vanishes is the transition point
($\tilde J_2 \simeq 1.193$),
the result shown in figure \ref{fig:width-1-3} being in very good agreement
with the above-mentioned result obtained by the LS method with the TBC.
We see a cusp in the curve at $\tilde J_2 \simeq 0.561$.
This cusp corresponds to the FRI-SF transition in figure \ref{fig:phase-m0}
(see also figure 19(b) below).
In fact, an infinitesimal small field brings about the $M=\Ms/3$ magnetization state
in the FRI region,
whereas a finite field is required for the $M=\Ms/3$ magnetization state
in the SF and D regions.

\begin{figure}[ht]
   \begin{center}
         \scalebox{0.4}[0.4]{\includegraphics{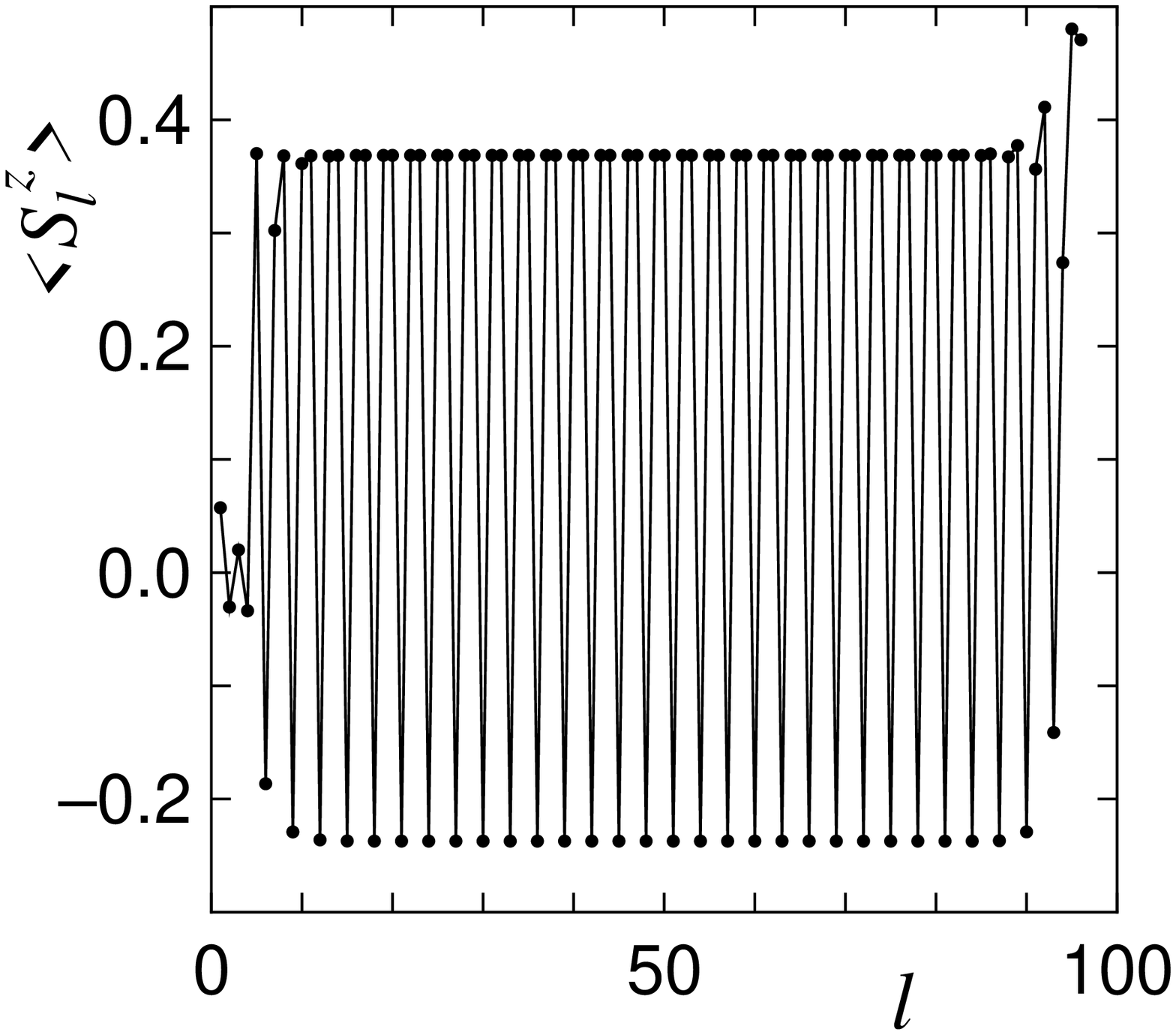}}
         \scalebox{0.4}[0.4]{\includegraphics{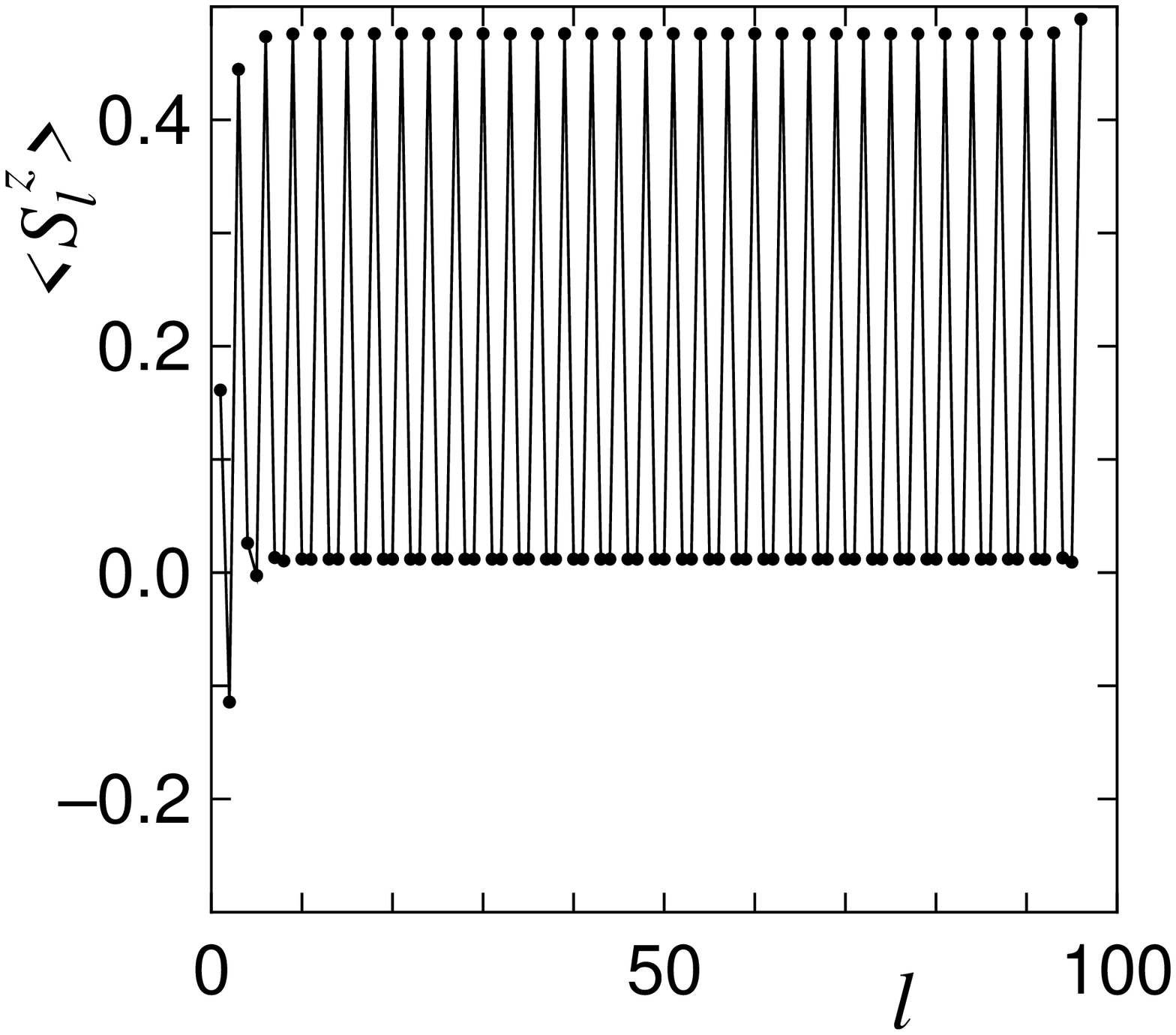}}
   \end{center}
   \caption{Behavior of $\ave{S_l^z}$ of the $N=96$ system (a) in the plateau
            A region, where $(\tilde J_2,\;\tilde J_3)=(0.80,\;0.55)$, and
           (b) in the plateau B region, where
           $(\tilde J_2,\;\tilde J_3)=(1.50,\;0.55)$.}
   \label{fig:config-1-3}
\end{figure}

Let us now discuss the expectation value $\ave{S_l^z}$ of each spin in the
$M=\Ms/3$ plateau state.  In the plateau A region, the approximate wave
function of the $j$th 3-spin cluster is
\begin{equation}
    \phi_{1,j} = 
    {1 \over \sqrt{6}}
    \left(   |\uparrow_{3j-1}\uparrow_{3j}\downarrow_{3j+1}\rangle
          - 2|\uparrow_{3j-1}\downarrow_{3j}\uparrow_{3j+1}\rangle
          +  |\downarrow_{3j-1}\uparrow_{3j}\uparrow_{3j+1}\rangle
    \right)
\end{equation}
which leads to
\begin{equation}
    \ave{S_{3j-1}^z} = \ave{S_{3j+1}^z} = {1 \over 3}~~~~~
    \ave{S_{3j}} = - {1 \over 6}.
\end{equation}
Thus, the expectation values $\ave{S_l^z}$ should be 
\begin{equation}
   \left\{ \cdots, \left( {1 \over 3}, -{1 \over 6}, {1 \over 3}\right),
          \left( {1 \over 3}, -{1 \over 6}, {1 \over 3}\right),
          \left( {1 \over 3}, -{1 \over 6}, {1 \over 3}\right),\cdots
   \right\}
\end{equation}
where three spins in a parenthesis form a 3-spin cluster.  In the plateau
B region, on the other hand, it is very easy to see that
$\ave{S_l^z}$ should be
\begin{equation}
   \left\{ \cdots, \left( 0, 0 \right),  {1 \over 2},
           \left( 0, 0 \right),  {1 \over 2},
           \left( 0, 0 \right),  {1 \over 2},
           \left( 0, 0 \right),  {1 \over 2},\cdots
   \right\}
\end{equation}
where two spins in a parenthesis form a singlet dimer.  Figure
\ref{fig:config-1-3} shows the DMRG result for $\ave{S_l^z}$ for the $N=96$
system in the plateau A region, where
$(\tilde J_2,\;\tilde J_3)=(0.80,\;0.55)$, and that in the plateau B region,
where $(\tilde J_2,\;\tilde J_3)=(1.50,\;0.55)$.  Then, the behavior of
$\langle S_l^z \rangle$ in this figure is quite consistent with our picture
of the mechanisms for the $M=\Ms/3$ plateau.

%********************************************
\section{Magnetization plateau at $M=(2/3)\Ms$}
%********************************************

Our incipient discussions on the $M=(2/3)\Ms$ plateau of the $S=1/2$ DD chain
model are given in a previous paper \cite{Tone-Oka-2}; here we aim at
discussing it in much more detail.  

Since the unit cell of the present model consists of three $S=1/2$ spins,
the OYA condition \cite{OYA} tells us that if the SBTS does not occur, the
possible magnetization plateau for $0<M<\Ms$ is only at $M=\Ms/3$.  Thus, the
SBTS is necessary for the realization of the $M=(2/3)\Ms$ plateau.  In the
3-spin cluster limit where $\tilde J_2=\tilde J_3=0$, the $M=(2/3)\Ms$
magnetization state is realized when a half of the 3-spin clusters are in the
$S_{\rm tot}^{(3)z}=1/2$ state and the remaining half are in the
$S_{\rm tot}^{(3)z}=3/2$ state.  The configuration of these two states are
completely free when $\tilde J_2=\tilde J_3=0$, resulting in the
$2^{N/3}$-fold degeneracy.  This high degeneracy is lifted when we introduce
$\tilde J_2$ and $\tilde J_3$.

\begin{figure}[ht]
   \begin{center}
         \scalebox{0.4}[0.4]{\includegraphics{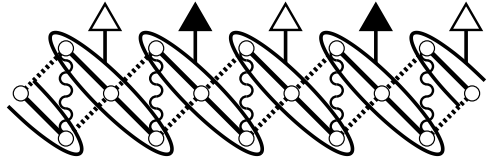}}
   \end{center}
   \caption{Physical picture of the $M=(2/3)\Ms$ plateau state for the
            $S=1/2$ DD chain model.  Ellipses with open and closed triangles
            denote the 3-spin clusters with the $\psi_1$ and $\psi_2$ states,
            respectively.} 
   \label{fig:picture-2-3}
\end{figure}

Let us discuss the $M=(2/3)\Ms$ magnetization plateau problem based on the
above picture by use of the degenerate perturbation theory
\cite{Totsuka}.  The lowest energy state of the $j$th 3-spin cluster with
$S_{\rm tot}^{(3)z}=1/2$ is given by
\begin{equation}
    \psi_{1,j}
    = {1 \over \sqrt{6}}
      \left(  |\uparrow_{3j-1}\uparrow_{3j}\downarrow_{3j+1}\rangle
            - 2|\uparrow_{3j-1}\downarrow_{3j}\uparrow_{3j+1}\rangle
            + |\downarrow_{3j-1}\uparrow_{3j}\uparrow_{3j+1}\rangle
      \right)
      \label{eq:psi1}
\end{equation}
which is nothing but equation (\ref{eq:sz+12}).  Here we use the symbol
$\psi_{1,j}$ for convenience.  The energy of this state is
\begin{equation}
    E_1 = - J_1 - {1 \over 2}H.
\end{equation}
On the other hand, the state of the $j$th 3-spin cluster with
$S_{\rm tot}^{(3)z}=3/2$ is
\begin{equation}
    \psi_{2,j}
    = |\uparrow_{3j-1}\uparrow_{3j}\uparrow_{3j+1}\rangle
      \label{eq:psi2}
\end{equation}
having the energy
\begin{equation}
    E_2 = {1 \over 2}J_1 - {3 \over 2}H.
\end{equation}
Using the pseudo-spin operator ${\bi T_j}$, where ${\bi T_j}^2=T_j(T_j+1)$
with $T_j=1/2$, we express the $\psi_{1,j}$ and $\psi_{2,j}$ states by the
$T_j^z=-1/2$ and $T_j^z = +1/2$ states, respectively.  We neglect the other 6
states of the $j$th 3-spin cluster, which can be justified near
$\tilde J_2=\tilde J_3 =0$.  We note that $E_1=E_2$ when the magnetic field
is given by $H=H_{\rm 2/3}^{(0)}\equiv(3/2)J_1$.  The lowest order
perturbation calculation with respect to $\tilde J_2$ and $\tilde J_3$
leads to the effective Hamiltonian for the pseudo-spin operator ${\bi T_j}$
\begin{equation}
   {\cal H}_{\rm eff}
   = \sum_j \{J^\perp_{\rm eff} (T_j^x T_{j+1}^x + T_j^y T_{j+1}^y)
           +J^z_{\rm eff} T_j^z T_{j+1}^z \}
     - H_{\rm eff} \sum_j T_j^z       
   \label{eq:Ham-T}
\end{equation}
where
\begin{eqnarray}
    &&J^\perp_{\rm eff} = {J_2 - 4J_3 \over 6} \\
    &&J^z_{\rm eff} = {J_2 + 8J_3 \over 36} \\
    &&H_{\rm eff} = H - H_{2/3}^{(0)} - {5J_2 + 22J_3 \over 36}.
\end{eqnarray}
The ground state of the effective Hamiltonian (\ref{eq:Ham-T}) is either
the N\'eel state or the SF state depending on whether $\Delta_{\rm eff}>1$
or $0\leq\Delta_{\rm eff}\le 1$, where $\Delta_{\rm eff}$ is 
\begin{equation}
    \Delta_{\rm eff}
    \equiv {J^z_{\rm eff} \over |J^\perp_{\rm eff}|}
    = {J_2 + 8 J_3 \over 6|J_2 - 4J_3|}.
\end{equation}
Thus, the N\'eel ground state is realized when
\begin{equation}
    {5 \over 32} < {J_3 \over J_2} < {7 \over 16}.
    \label{eq:boundary-2-3}
\end{equation}
The N\'eel and SF ground states in the ${\bi T}$-picture
correspond to the $M=(2/3)\Ms$ plateauful and plateauless states in the
original ${\bi S}$-picture, respectively.  The transition between the
plateauful and plateauless states is of the BKT type
\cite{Berezinskii,KT}.  The magnetic field $H_{2/3}$ given by the solution of
$H_{\rm eff}=0$, i.e.,
\begin{equation}
    H_{2/3}
    \equiv H_{2/3}^{(0)} + {5J_2 + 22J_3 \over 36} 
    = {54J_1+ 5J_2 + 22J_3 \over 36}.
\end{equation}
represents the magnetic field corresponding to the center of the magnetization
plateau in the plateauful case, and that corresponding to the $M=\Ms/3$
magnetization in the plateauless case.  This degenerate perturbation
calculation has already been developed in
\cite{Tone-Oka-2,Honecker-Lauchli}.  The physical picture of the
$M=(2/3)M_{\rm }$ plateau is shown in figure \ref{fig:picture-2-3}.

For the purpose of drawing the $M=(2/3)\Ms$ plateau phase diagram,
we have performed
the numerical diagonalization of the finite size Hamiltonian by the Lanczos
method.  We can apply the LS method \cite{Oka-Nom,Oka-LS,Nom-LS,Nom-Oka,Nom}
to this kind of plateauful-plateauless transition of the BKT type
\cite{Okazaki-Okamoto-Sakai,Okamoto-Okazaki-Sakai-1,Nakasu,Okamoto-Okazaki-Sakai-2} accompanied by the SBTS.  The plateauful-plateauless transition point can
be obtained from the crossing point of $\Delta E_0(N,M=(2/3)\Ms)$ and
$\Delta E_{\pm 1}(N,M=(2/3)\Ms)$ as functions of the quantum
parameters.  Here, $\Delta E_0(N,M)$ and $\Delta E_{\pm 1}(N,M)$ are defined,
respectively, by
\begin{eqnarray}
    &&\Delta E_0(N,M)
    \equiv E_1\left(N, M \right)
          -E_0\left(N, M \right) \\
    &&\Delta E_{\pm 1}(N,M)
    \equiv {1 \over 2} 
          \left\{ E_0\left(N, M + 1\right)
                + E_0\left(N, M - 1 \right)
          \right\}
          - E_0\left(N, M \right)
\end{eqnarray}
where $E_0(N,M)$ and $E_1(N,M)$ are, respectively, the lowest and second lowest
energies in the subspace of the magnetization $M$ for the finite size system
with $N$ spins.  It is noted that the $M=(2/3)\Ms$ state is either plateauful
(the N\'eel state in the ${\bi T}$-picture) or plateauless (the SF state in
the ${\bi T}$-picture) depending on whether
\hbox{$\Delta E_0(N,M=(2/3)\Ms)<\Delta E_{\pm 1}(N,M=(2/3)\Ms)$} or
\hbox{$\Delta E_0(N,M=(2/3)\Ms)>\Delta E_{\pm 1}(N,M=(2/3)\Ms)$}.

\begin{figure}[ht]
   \begin{center}
         \scalebox{0.4}[0.4]{\includegraphics{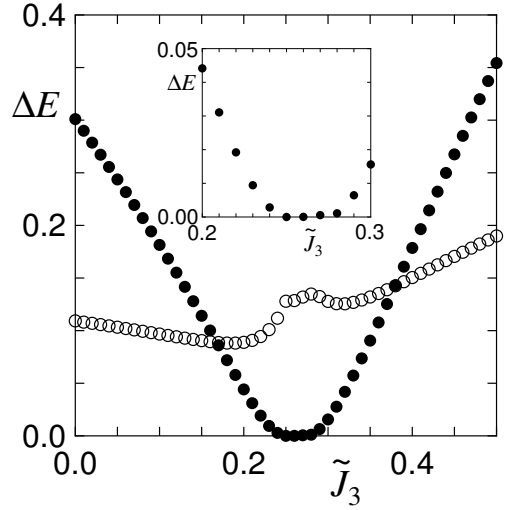}}
   \end{center}
   \caption{$\Delta E_0(N,M=(2/3)\Ms)$ (closed circles) and 
            $\Delta E_{\pm 1}(N,M=(2/3)\Ms)$ (open circles)
            with $N=18$ ($\Ms=9$) as functions of $\tilde J_3$ when
            $\tilde J_2=0.80$.  
            Approximated values for the plateauful-plateauless
            phase boundary points in the $N \to \infty$ limit
            are known from the crossings of $\Delta E_0(N,M=(2/3)\Ms)$ and 
            $\Delta E_{\pm 1}(N,M=(2/3)\Ms)$. 
            The point where $\Delta E_0(N,M=(2/3)\Ms)$ vanishes
            ($\tilde J_3 \simeq 0.25$; see the inset)
            corresponds to the Ising case
            of the pseudo-spin Hamiltonian (\ref{eq:Ham-T}).}
   \label{fig:cross(2-3)}
\end{figure}

\begin{figure}[ht]
   \begin{center}
         \scalebox{0.4}[0.4]{\includegraphics{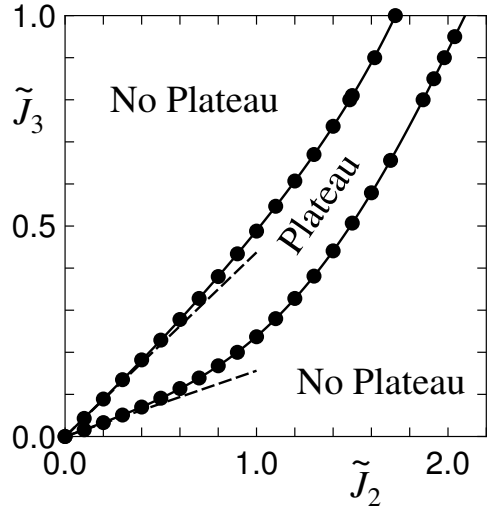}}
   \end{center}
   \caption{$M=(2/3)\Ms$ plateau phase diagram of the $S=1/2$ DD chain
            model.  Closed circles denote the numerical results,
            and the dashed lines the results of the degenerate 
            perturbation calculation given by equation
            (\ref{eq:boundary-2-3}).}
   \label{fig:phase-2-3}
\end{figure}

Figure \ref{fig:cross(2-3)} shows the behaviors of $\Delta E_0(N,M=(2/3)\Ms)$
and $\Delta E_{\pm 1}(N,M=(2/3)\Ms)$ with $N=18$ ($\Ms=9$) as functions of
$\tilde J_3$ when $\tilde J_2=0.80$.
The value of $\tilde J_3$ at each
crossing point is expected to be an approximate value for the
plateauful-plateauless transition point $\tilde J_3^{\rm (cr)}$ in the
$N\to\infty$ limit.  We have carried out, to estimate $\tilde J_3^{\rm (cr)}$,
an extrapolation to this limit similar to that in the case of the estimation
of $\tilde J_2^{\rm (cr)}$ for the $M=\Ms/3$ plateau phase diagram, discussed
in \S3, and have obtained the $M=(2/3)\Ms$ plateau phase diagram depicted in
figure \ref{fig:phase-2-3}.
In this figure a good agreement is found
between the numerical result and the analytical prediction given by equation
(\ref{eq:boundary-2-3}) near $(\tilde J_2,\tilde J_3)=(0,0)$.
We note that the point $\Delta E_0 = 0$ ($\tilde J_3 \simeq 0.25$)
corresponds to the Ising case
($J_{\rm eff}^\perp = 0$) of the pseudo-spin Hamiltonian (\ref{eq:Ham-T}).
The double degeneracy in the N\'eel ground state is realized only in the
$N \to \infty$ limit in usual cases,
whereas it is realized even in finite size systems for the Ising case.
Thus, the point $\Delta E_0 = 0$, which means the doubly degenerate ground state,
corresponds to the Ising case.

\begin{figure}[ht]
   \begin{center}
         \scalebox{0.4}[0.4]{\includegraphics{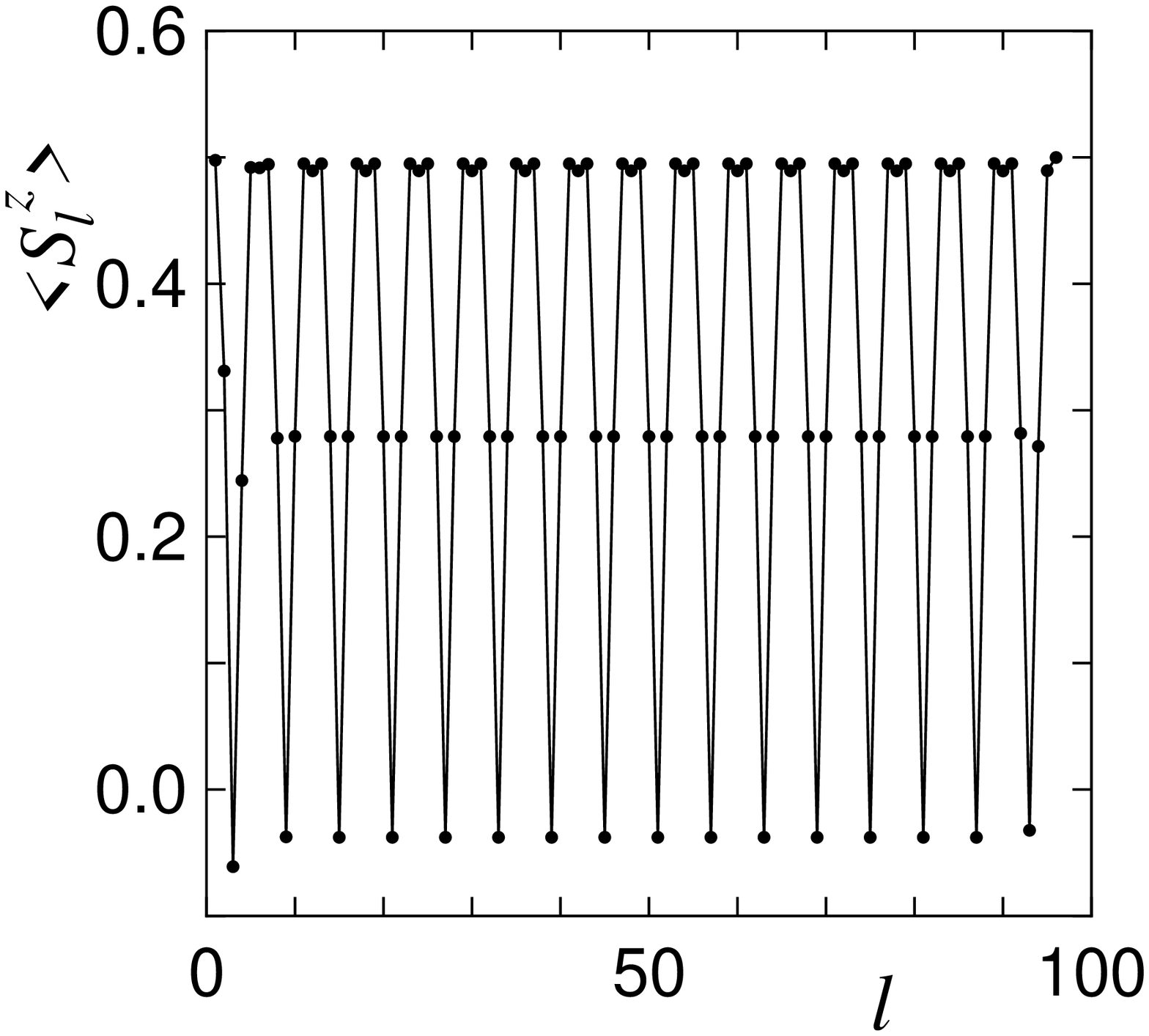}}
         \scalebox{0.4}[0.4]{\includegraphics{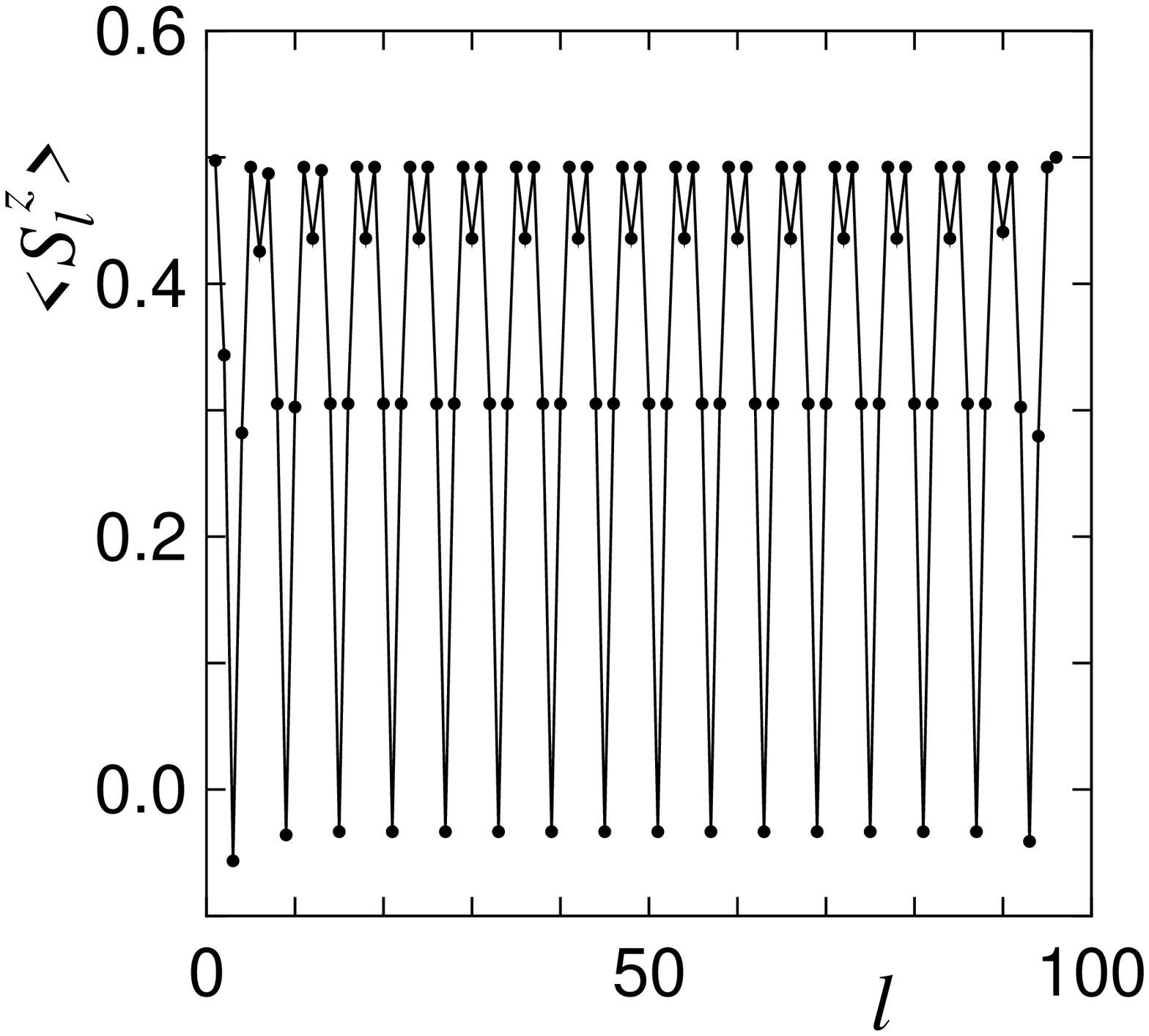}}
   \end{center}
   \caption{Behavior of $\langle S_l^z \rangle$ of the $N=96$ system in
    the $M=(2/3)\Ms$ plateau state, where
    (a) $(\tilde J_2,\;\tilde J_3)=(0.80,\;0.27)$
    and 
    (b) $(\tilde J_2,\;\tilde J_3)=(0.80,\;0.30)$
    }
   \label{fig:config-2-3}
\end{figure}

The expectation value $\langle S_l^z\rangle$ of the $(2/3)\Ms$
plateau state can be obtained from the above-mentioned physical picture
(figure \ref{fig:picture-2-3}); equations (\ref{eq:psi1}) and (\ref{eq:psi2})
yields for $\langle S_l^z\rangle$
\begin{equation}
   \left\{ \cdots, 
          \overbrace{\left( {1 \over 3}, -{1 \over 6}, 
          {1 \over 3}\right)}^{\displaystyle{\psi_1}},
          \overbrace{\left( {1 \over 2}, {1 \over 2}, 
          {1 \over 2}\right)}^{\displaystyle{\psi_2}},
          \overbrace{\left( {1 \over 3}, -{1 \over 6}, 
          {1 \over 3}\right)}^{\displaystyle{\psi_1}},
          \overbrace{\left( {1 \over 2}, {1 \over 2}, 
          {1 \over 2}\right)}^{\displaystyle{\psi_2}}, 
          \cdots
   \right\}
   \label{eq:config-2-3}
\end{equation}
where three spins in a parenthesis form a 3-spin cluster.
Figure \ref{fig:config-2-3} shows the behaviors of $\langle S_l^z\rangle$
which are obtained by using the DMRG method for the $N=96$ system
in the $(\tilde J_2,\;\tilde J_3)=(0.80,\;0.27)$
and $(\tilde J_2,\;\tilde J_3)=(0.80,\;0.30)$ cases.
As can be readily seen, the
behavior of $\langle S_l^z\rangle$ is quite consistent with equation
(\ref{eq:config-2-3}).
Of course, the ideal behavior given by this equation
is expected in the $\Delta_{\rm eff}=\infty$ limit (the Ising limit) in
the effective Hamiltonian (\ref{eq:Ham-T}), even if the mapping onto the
${\bi T}$-picture is justified.
In the case where $\Delta_{\rm eff}$ is
finite, the hybridization between the $\psi_1$ and $\psi_2$ states occurs,
which well explains the fact that
$\ave{S_{3j}^z}<\ave{S_{3j-1}^z}=\ave{S_{3j+1}^z}$ in the $\psi_2$-rich
cluster. 
The behavior of $\langle S_l^z\rangle$ of the $\psi_2$-rich cluster 
in the $(\tilde J_2,\;\tilde J_3)=(0.80,\;0.27)$ case
is nearer to the ideal behavior than that in the
$(\tilde J_2,\;\tilde J_3)=(0.80,\;0.30)$ case.
This suggests that the parameter set in the former case is nearer to the 
Ising limit of the effective Hamiltonian (\ref{eq:Ham-T}) 
than that in the latter case.
In fact, the Ising limit is realized when $\tilde J_3 \simeq 0.25$
for $\tilde J_2 = 0.55$, 
as can be seen from figure 15.

%********************************************
\section{Magnetization curves and magnetization phase diagrams}
%********************************************

In order to calculate the ground state magnetization curve of the present
$S=1/2$ DD chain model, we have employed the DMRG method
\cite{White1,White2}.  The procedure of our DMRG calculation is briefly
summarized in reference \cite{Tone-Oka-1}.

We have computed $E_0(N,M)$ for finite size systems with up to $N=96$
($N=216$) spins for general (special) values of $\tilde J_2$ and
$\tilde J_3$.  Once the values of $E_0(N,M)$ for all nonnegative $M$'s ($M=0$,
$1$, $2$, $\cdots$, $N/2$) are known, the ground state magnetization curve
can be obtained by plotting as a function of $H$ the average magnetization
$\langle m \rangle$ per one spin, where $\langle m \rangle$ is the value of
$m$ which yields the minimum among $\{E_0(N,mN)/N-Hm\}$'s with $m=0$, $1/N$,
$2/N$, $\cdots$, $1/2$.  As is naturally expected, the resulting magnetization
curve is a stepwisely increasing function of $H$.  Following Bonner and
Fisher's pioneering work \cite{BF}, we may obtain, except for plateau regions,
a satisfactorily good approximation to the ground state magnetization curve
in the $N\to\infty$ limit by drawing a smooth curve through the midpoints of
the steps in the finite size results.  As for the plateau regions, on the
other hand, we can estimate the lowest and highest values of $H$ in the
$N\to\infty$ limit, denoted by $H_{p,{\rm l}}$ and $H_{p,{\rm h}}$,
respectively, giving the $M=p\Ms$ ($p=0$, $1/3$, $2/3$ or $1$) plateau by
extrapolating the corresponding finite size values to this limit.

\begin{figure}[ht]
   \begin{center}
         \scalebox{0.4}[0.4]{\includegraphics{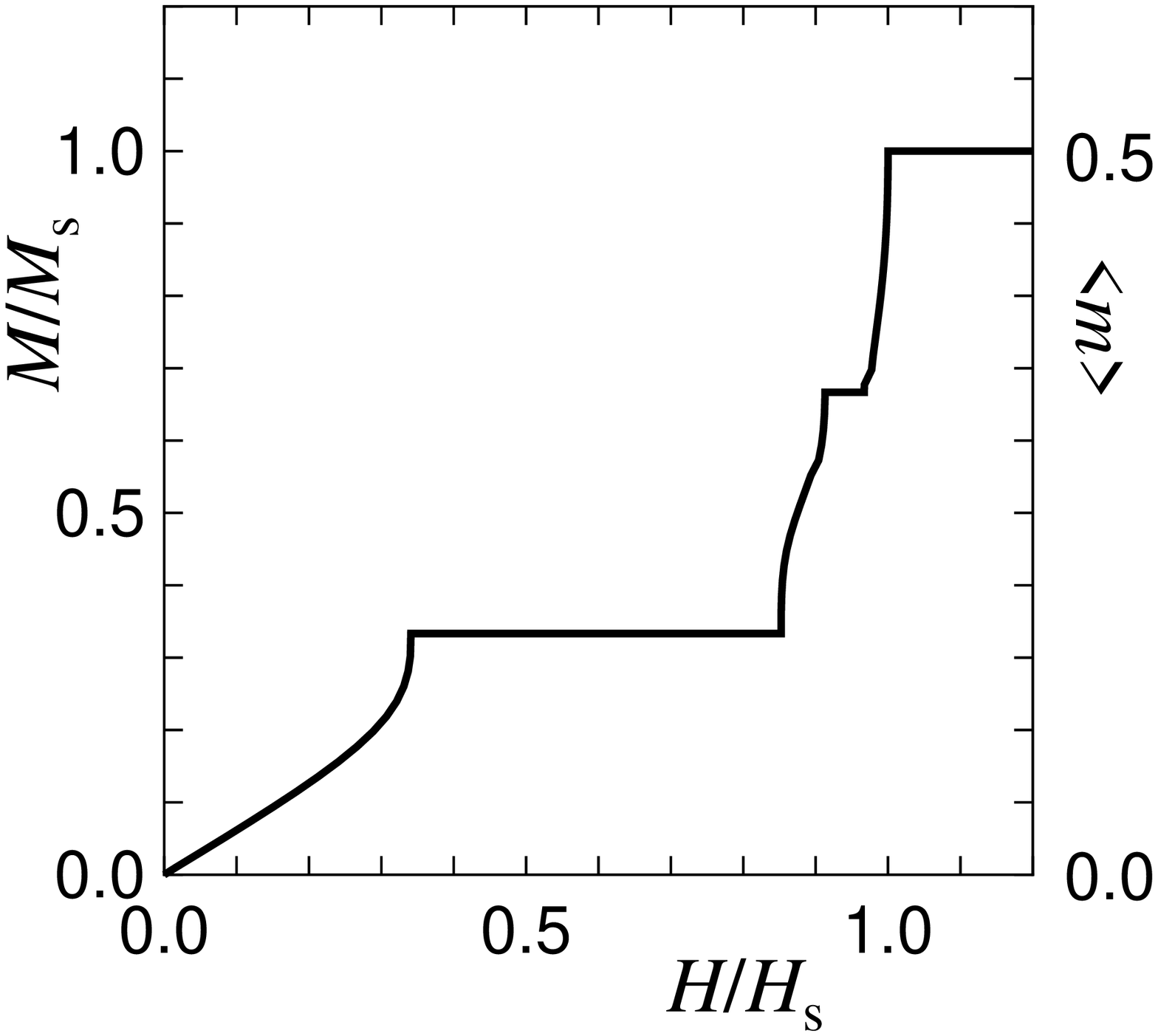}}~~~~~~~
         \scalebox{0.4}[0.4]{\includegraphics{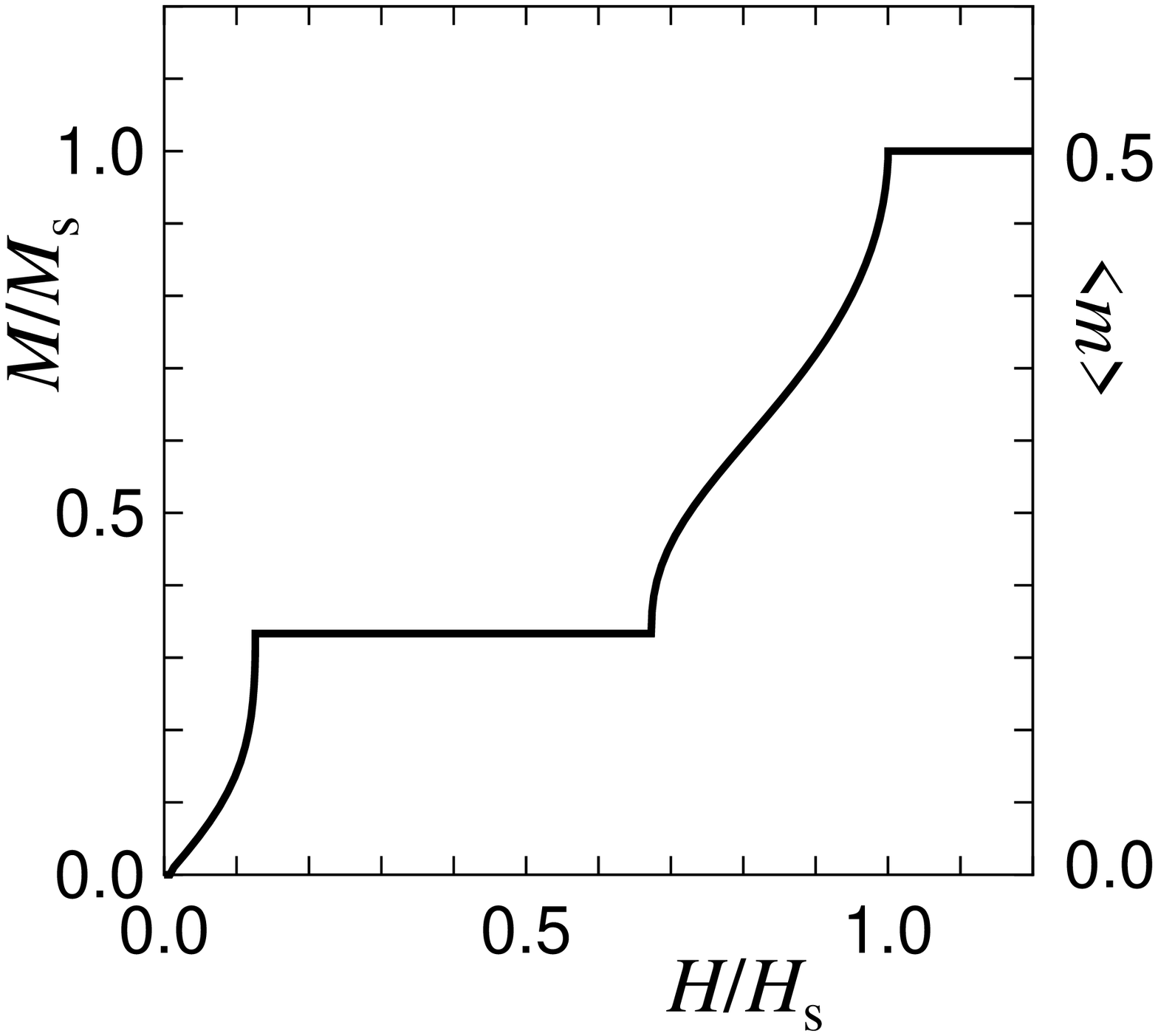}}
   \end{center}
   \caption{Magnetization curves in the $N\to\infty$ limit for (a)
            $(\tilde J_2,\;\tilde J_3)=(0.80,0.27)$ and (b)
            $(\tilde J_2,\;\tilde J_3)=(0.80,0.55)$.  Note that $\Hs$ denotes
            the saturation field.  The wide plateau at $M=\Ms/3$ is observed
            in both (a) and (b).  The $M=(2/3)\Ms$ plateau can be seen in (a),
            the parameters for which lies in the plateau region shown in
            figure \ref{fig:phase-2-3}.}
   \label{fig:mag-curve}
\end{figure}

\begin{figure}[ht]
   \begin{center}
         \scalebox{0.4}[0.4]{\includegraphics{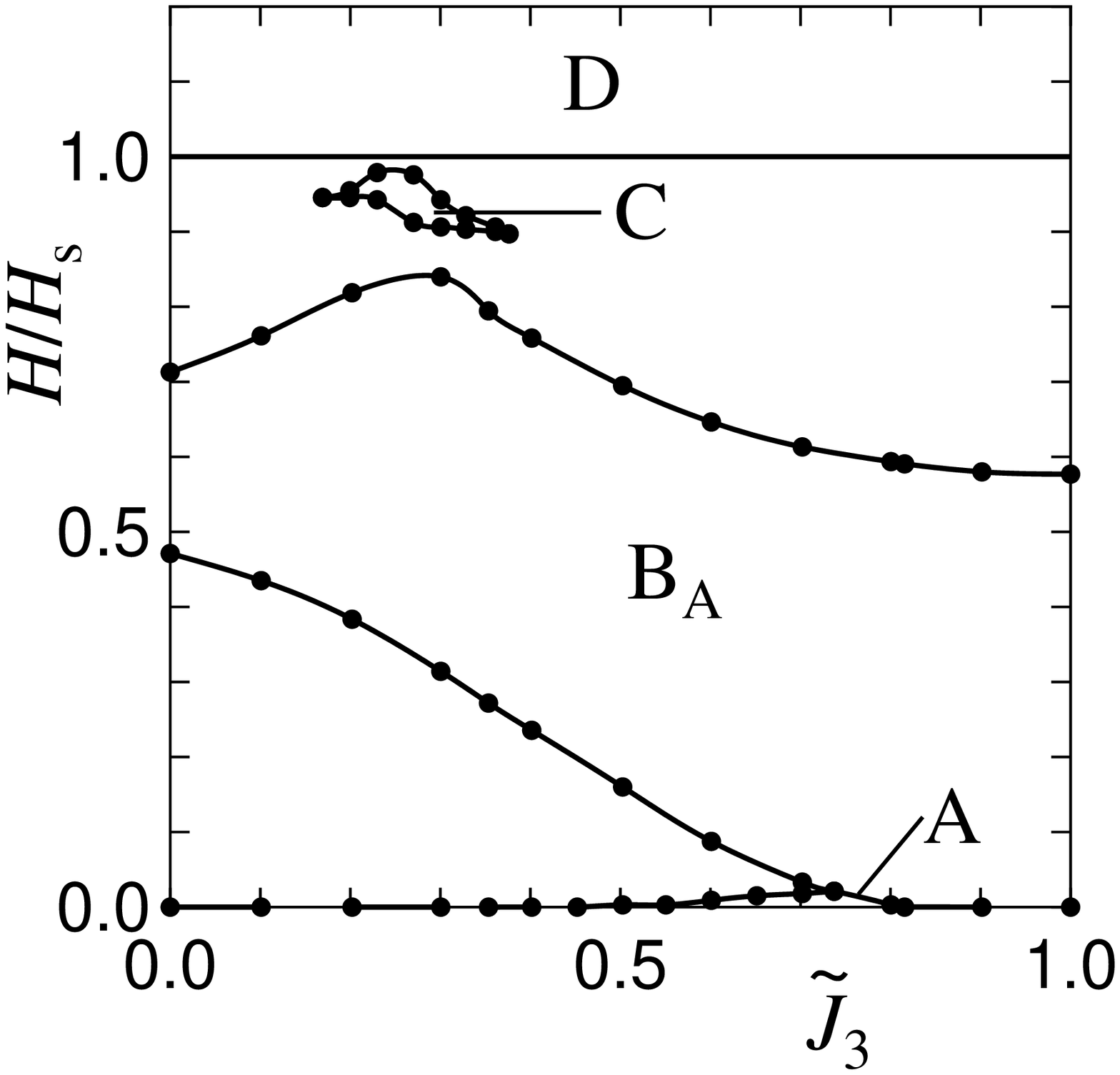}}~~~~~
         \scalebox{0.4}[0.4]{\includegraphics{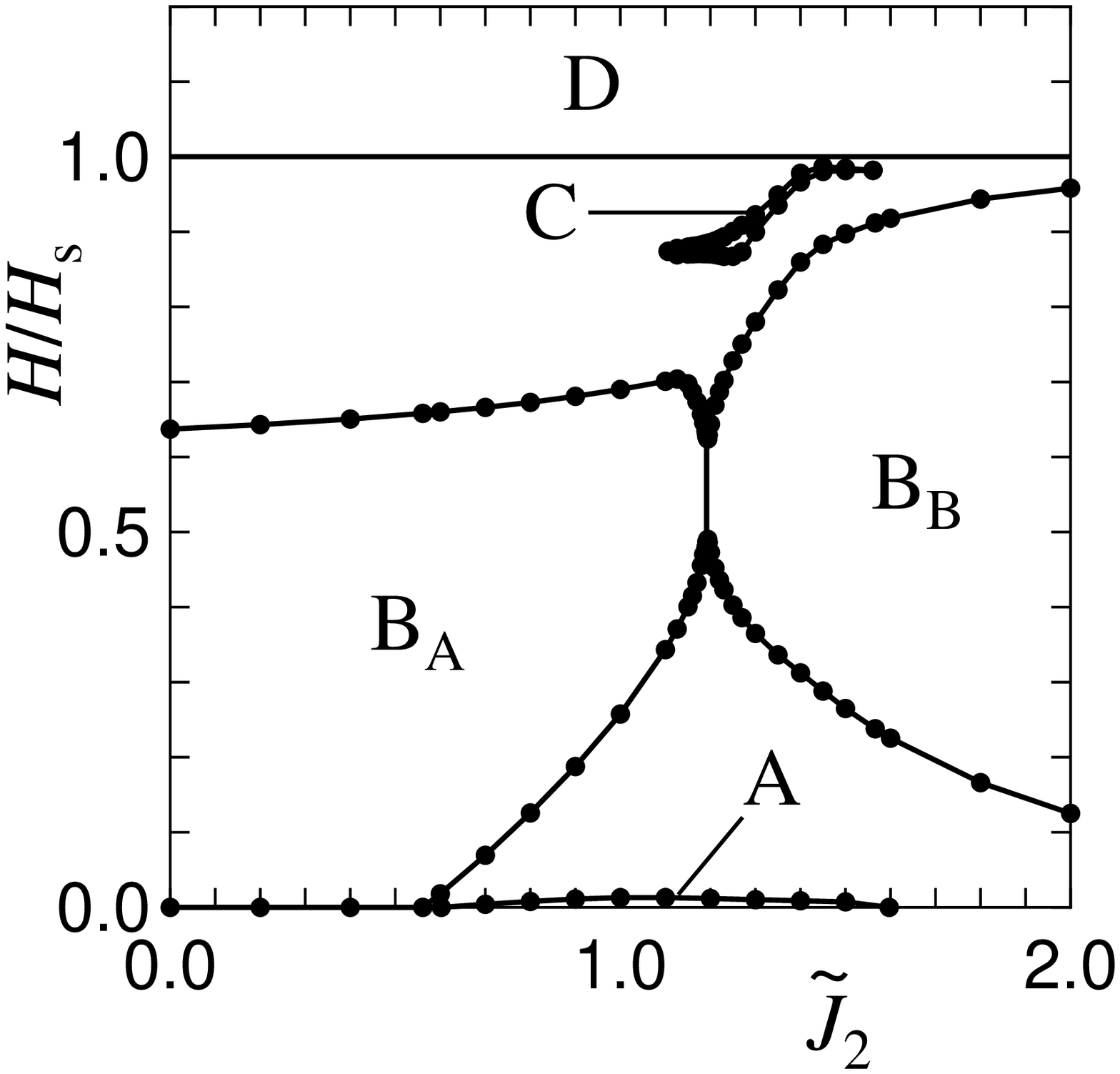}}
   \end{center}
   \caption{Magnetization phase diagrams (a) on the $H/H_{\rm s}$
            versus $\tilde J_3$ plane when $\tilde J_2 = 0.80$
            and (b) on the $H/H_{\rm s}$ versus $\tilde J_2$ plane when
            $\tilde J_3=0.55$.  In the regions A, $\rm{B_X}$
            (${\rm X=A,B}$), C and D, the average magnetization
            $\langle m \rangle$ per one spin is equal to $0$ (the $M=0$
            plateau region), $1/6$ (the $M=\Ms/3$ plateau region), $1/3$ (the
            $M=(2/3)\Ms$ plateau region, and 1 (the $M=\Ms$ plateau, or the
            saturated magnetization, region, respectively; the B$_{\rm A}$
            and B$_{\rm B}$ regions are, respectively, for the plateaux A
            and B.  In the other regions, $\langle m \rangle$ varies
            continuously with the increase of $H$.  For (b), we can see the
            closing of the $M=\Ms/3$ plateau at $\tilde J_2=1.193$, which
            corresponds to the phase transition between the plateau A and B
            phases, as expected from the $M=\Ms/3$ phase diagram shown in
            figure \ref{fig:phase-1-3}.  For (a), such a transition is not
            seen, which is also consistent with the $M=\Ms/3$ phase diagram.}
   \label{fig:mag-phase}
\end{figure}

The magnetization curves in the $N\to\infty$ limit thus obtained for
the $(\tilde J_2,\;\tilde J_3)=(0.80,\;0.27)$ case and the
$(\tilde J_2,\;\tilde J_3)=(0.80,\;0.55)$ case are shown in figure
\ref{fig:mag-curve}.  It is noted that $H_{p,{\rm l}}$ and
$H_{p,{\rm h}}$ in both cases are estimated by fitting the finite size
results for $N=48$, $72$ and $96$ to a quadratic function of $1/N^2$, which
is similar to equation (\ref{eq:extrapolation}).  The $H=0$ ground state phase
in the former case is the SF phase (see figure \ref{fig:phase-m0}), and
therefore we have no $M=0$ plateau.  Since that in the latter case is the D
phase, on the other hand, we have in principle a finite $M=0$ plateau,
although the $M=0$ plateau width is so narrow that we cannot see it
clearly (see figure \ref{fig:mag-phase}(b) below).  The wide $M=\Ms/3$
plateaux are found in both cases, while the $M=(2/3)\Ms$ plateau is observed
only in the former case, which is consistent with the $M=(2/3)\Ms$ plateau
phase diagram shown in figure \ref{fig:phase-2-3}.

We have also obtained the magnetization phase diagrams on the
$(\tilde J_3,\;H)$ plane in the case of $\tilde J_2 =0.80$
and on the $(\tilde J_2,\;H)$ plane in the case of $\tilde J_3=0.55$.
The results are depicted in figure \ref{fig:mag-phase}, where
$H_{0,{\rm h}}/\Hs$, $H_{1/3,{\rm l}}/\Hs$, $H_{1/3,{\rm h}}/\Hs$,
$H_{2/3,{\rm l}}/\Hs$, $H_{2/3,{\rm h}}/\Hs$ as well as $H/\Hs=1.0$, where
$\Hs(\equiv H_{1,{\rm l}})$ is the saturation field, are plotted.  When we
estimate these values for general values of $\tilde J_2$ and $\tilde J_3$,
we have employed the same method as that for the estimations of
$\tilde J_2^{\rm (cr)}$ in \S3 and $\tilde J_3^{\rm (cr)}$ in \S4.  It should
be emphasized, however, that in estimating $H_{1/3,{\rm l}}$ and
$H_{1/3,{\rm h}}$ for special values of $\tilde J_2=1.190$, $1.191$, $1.192$,
$1.194$ and $1.195$ with $\tilde J_3=0.55$, we have used the finite size
results for $N=132$, $168$ and $216$ instead of those for $N=48$, $72$ and
$96$, since the $N$ dependence of the finite size results are rather strong
for these $\tilde J_2$ and $\tilde J_3$.  In the case of $\tilde J_2 =0.80$
the change of the mechanism of the $M=\Ms/3$ plateau is not observed, whereas
in the case of $\tilde J_3=0.55$, it is clearly observed.  This is quite
consistent with the $M=\Ms/3$ plateau phase diagram shown in figure
\ref{fig:phase-2-3}.  We note that in the case of $\tilde J_3=0.55$, the
$M=\Ms/3$ plateau starts from $H=0$ when $0\leq\tilde J_2<0.561$, since the
$H=0$ ground state phase is the FRI phase.  Finally, we mention that the
$M=\Ms/3$ plateau width $W_{(1/3)}$ plotted in figure \ref{fig:width-1-3} is
defined by \begin{equation}
   W_{(1/3)} = H_{1/3,{\rm h}} - H_{1/3,{\rm l}}.
   \label{eq:W-1-3}
\end{equation}

%********************************************
\section{Discussion and Summary}
%********************************************

We have investigated the magnetic properties of the $S=1/2$ DD chain model
at $T=0$ by use of the degenerate perturbation theory, the LS analysis of the
numerical diagonalization data obtained by the Lanczos method, and the DMRG
calculation.  We have clarified the mechanism for the $M=\Ms/3$ and
$M=(2/3)\Ms$ plateaux, and have also drawn the precise plateau phase diagrams
by carrying out the LS analysis.  These phase diagrams agree with the results
obtained by physical considerations and the degenerate perturbation
theory.  The expectation value $\ave{S_l^z}$ of each spin estimated by the
DMRG calculation strongly supports the plateau formation mechanisms.  The
magnetization curves for a few sets of the parameters $\tilde J_2$ and
$\tilde J_3$ have been obtained by the DMRG calculation, the behavior of which
are consistent with the plateau phase diagrams.  We have also drawn the
magnetization phase diagram on the $(\tilde J_3,\;H)$ plane for
$\tilde J_2=0.80$ and that on the $(\tilde J_2,\;H)$ plane for
$\tilde J_3=0.55$.

As noted in \S1, 
some substances are thought to be well modeled by the DD chain.
${\rm A_3 Cu_3 (PO_4)_4}$ with ${\rm A = Ca, \,Sr}$
were investigated by Drillon et al. \cite{Drillon1,Drillon2}
and Ajiro et al. \cite{Ajiro}.
The measured magnetic susceptibility $\chi(T)$ \cite{Drillon1,Drillon2}
showed the ferrimagnetic behavior.
In the high field magnetization process measurement by
Ajiro et al. \cite{Ajiro},
the magnetization rapidly reached the wide $M=\Ms/3$ plateau
as the magnetic field was increased.
The $M=\Ms/3$ plateau was kept even at the
high field limit 40T in their experiment.
Ajiro et al. \cite{Ajiro} performed the neutron scattering experiment
for ${\rm Sr_3 Cu_3 (PO_4)_4}$ at $T=4\,{\rm K}$ and $T=70\,{\rm mK}$.
The transition temperature of the
three-dimensional ordering to the antiferromagnetic state
of ${\rm Sr_3 Cu_3 (PO_4)_4}$ is $T_{\rm N}=0.96\,{\rm K}$.
Although their neutron diffraction pattern at $T=70\,{\rm mK}$
was consistent with the antiferromagnetic ordering,
its magnitude was strongly reduced to almost $1/10$ of that in usual cases.
This fact suggests that the magnitude of spins on the average is about
$1/\sqrt{10} \sim 1/3$ of that for independent spins,
which is consistent with the picture of the 3-spin cluster formation
with $S_{\rm tot}^{(3)z} = \pm 1/2$.
Thus, we think, the parameter set of this substance
lies in the FRI region on the phase diagram figure \ref{fig:phase-m0}.

Sakurai et al. \cite{Sakurai}
investigated the magnetic susceptibility, magnetization process (up to 28T)
and ${\rm {}^{51}V}$ NMR of ${\rm Bi_4 Cu_3 V_2 O_{14}}$.
Unfortunately, they could not obtain any conclusion 
on the existence of the $M=\Ms/3$ plateau,
because $M \simeq 0.27\Ms$ at $H=28\,{\rm T}$.
This suggests that the ground state of this substance above the 
three dimensional antiferromagnetic ordering temperature
($T_{\rm N}=6\,{\rm K}$) is the SF state.
The ferrimagnetic behavior was not seen in the susceptibility
above $T_{\rm N}$, 
which is consistent with the SF ground state.

Kikuchi et al. \cite{Kikuchi1, Kikuchi2, Kikuchi3} have
investigated experimentally the magnetic and thermal properties of azurite
$\rm Cu_3(OH)_2(CO_3)_2$, which seems to be modeled by the $S=1/2$ DD chain
model.  Their result \cite{Kikuchi1} shows that the $H=0$ ground state of
azurite is the SF state.  Furthermore, their recent result \cite{Kikuchi3}
for the low temperature magnetization curve obtained by applying the magnetic
field along the $b$ axis (the chain axis) demonstrates that the $M=\Ms/3$
plateau appears in the field range of $0.5\Hs\lsim H\lsim0.8\Hs$ with
$\Hs\sim32.5 {\rm T}$.  These two results may imply that $\tilde J_3$ is
considerably smaller than unity and $\tilde J_2$ is considerably
larger than unity, 
as can be seen from figure \ref{fig:phase-m0}, since the $M=\Ms/3$ plateau width
should be much wider if $\tilde J_3\sim 1$ and $\tilde J_2\gsim 2$. 
It is noted that the fact that $\tilde J_3$ is smaller than unity is not
inconsistent with the fact \cite{Azurite} that the length of the bond between
the $(3j-1)$th and $(3j)$th sites is almost equal to that
of the bond between the $(3j)$th and $(3j+1)$th sites,
since the superexchange paths
connecting ${\bi S}_{3j-1}$ and ${\bi S}_{3j}$
and that connecting ${\bi S}_{3j}$ and ${\bi S}_{3j+1}$
are different with each other.  
Thus, we conjecture that the $M=\Ms/3$ plateau observed in
azurite is the plateau B shown in figure \ref{fig:pl-mech}(b) (see figure
\ref{fig:phase-1-3}). 
In addition to the above experimental results, Kikuchi
et al. \cite{Kikuchi3} have also found a plateau-like behavior at $M=(2/3)\Ms$
in the $dM/dH$ versus $H$ curve.
This may suggest that the value of
$\tilde J_2$ is not so much large compared with unity (see figure
\ref{fig:phase-2-3}).
Furthermore, they \cite{Kikuchi3} have observed the
anisotropy concerning the direction of the applied magnetic field in the
low temperature magnetization curves.
Thus, in order to quantitatively
explain the magnetic properties of azurite, more detailed investigations,
which take at least the antisymmetric Dzyaloshinsky-Moriya interactions
\cite{Dzyalo,Moriya} into account, are definitely required.  We, however,
believe that we have succeeded in explaining them at least qualitatively.

The specific heat measurement of azurite has also been done by Kikuchi
et al. \cite{Kikuchi3}.  They have found two peaks in the specific heat
$C(T)$ versus temperature $T$ curve, which suggests the existence of two
characteristic energies.  We consider that the larger one is $J_2$ and the
other is $(J_1-J_3)^2/2J_2$ which appears in equation (\ref{eq:Jeff}).

\section*{Acknowledgements}

We would like to express our appreciation to H.~Kikuchi for informing us
of their experimental results for azurite prior to publication and for
invaluable discussions.
We are deeply grateful to Y.~Ajiro
for information and useful discussions on 
${\rm A_3 Cu_3 (PO_4)_4}$ with ${\rm A = Ca, \,Sr}$.
We also thank T.~Hikihara by whom the DMRG
program employed in this study is coded,
and H.~Nishimori for the numerical diagonalization program
package TITPACK Ver.2.
Thanks are further due to
H.-J.~Mikeska, K.~Takano, K.~Kubo, H.~Tanaka, H.~Ohta, K.~Nomura,
A.~Kitazawa and A.~Honecker for stimulating discussions. 
This work has been partly supported by a
Grant-in-Aid for Scientific Research on Priority Areas (B)
(\lq\lq Field-Induced New Quantum Phenomena in Magnetic Systems\rq\rq)
and a Grant-in-Aid for Scientific Research (C) (No.~14540329)
from the Ministry of Education, Culture, Sports, 
Science and Technology of Japan. 
Finally, we
thank the Supercomputer Center, Institute for Solid State Physics, University
of Tokyo, the Information Synergy Center, Tohoku University and the Computer
Room, Yukawa Institute for Theoretical Physics, Kyoto University for
computational facilities.

%**********************
\end{document}